\documentclass[journal]{IEEEtran}

\usepackage{amssymb}
\usepackage{multirow}
\usepackage{balance}  % Add in the preamble
\usepackage{flushend}

\usepackage{cite}
\usepackage{url}
\usepackage{xcolor} 
\usepackage{comment}
\usepackage{color}
\usepackage{algorithm}         % To add algorithm environment
\usepackage{algpseudocode}     % To get the algorithmic commands
\usepackage{xspace}
\usepackage{amsmath}
\usepackage{graphicx}
\usepackage{subcaption}
\usepackage{makecell} 
\usepackage{booktabs} 
\usepackage{enumitem}
\usepackage{pifont}
\usepackage{placeins}
\usepackage{amssymb}
\usepackage{multirow}
\usepackage[bottom]{footmisc}

\usepackage{caption}
\usepackage{stfloats}   
\usepackage{afterpage}

\newcommand{\pname}{\texttt{xDiff}\xspace}

\usepackage{enumitem}
\setlist[itemize]{leftmargin=*}

\newcommand{\mycut}[1]{\textcolor{cyan}{}}

\begin{document}
% \settopmatter{printacmref=false, printfolios=false}
% \pagestyle{plain}
\title{
xDiff: Online Diffusion Model for Collaborative Inter-Cell Interference Management in 5G O-RAN}

\author{Peihao Yan\IEEEauthorrefmark{1}, 
Huacheng Zeng\IEEEauthorrefmark{1},
and
Y. Thomas Hou\IEEEauthorrefmark{2}\\
\IEEEauthorrefmark{1}Department of Computer Science and Engineering, Michigan State University\\
\IEEEauthorrefmark{2}Department of Electrical and Computer Engineering, Virginia Tech
}

\maketitle
\begin{abstract}

Open Radio Access Network (O-RAN) is a key architectural paradigm for 5G and beyond cellular networks, enabling the adoption of intelligent and efficient resource management solutions. 
% . In the O-RAN framework, online learning has emerged as a crucial technique for near-real-time (Near-RT) resource management and admission control via xAPPs hosted in the Near-RT RAN Intelligent Controller (RIC). 
Meanwhile, diffusion models have demonstrated remarkable capabilities in image and video generation, making them attractive for network optimization tasks. 
In this paper, we propose \pname, a diffusion-based reinforcement learning (RL) framework for inter-cell interference management (ICIM) in O-RAN. 
We first formulate ICIM as a resource allocation optimization problem aimed at maximizing a user-defined reward function and then develop an online learning solution by integrating a diffusion model into an RL framework for near-real-time policy generation. 
% To address the time-scale discrepancy between RIC and RAN operations, 
Particularly, we introduce a novel metric, \textit{preference values}, as the policy representation to enable efficient policy-guided resource allocation within O-RAN distributed units (DUs). 
We implement \pname on a 5G testbed consisting of three cells and a set of smartphones in two small-cell scenarios. Experimental results demonstrate that \pname outperforms state-of-the-art ICIM approaches, highlighting the potential of diffusion models for online optimization of O-RAN.
Source code is available on GitHub \cite{O-RANICIMPeihao}.

\end{abstract}


% \keywords{O-RAN, inter-cell interference management, xApp, Near-Real-Time RIC, conditional diffusion policy, online learning}



\section{Introduction}
%Conventional Gaussian policies may fail to fit datasets with complex distributions due to their restricted expressiveness.

Open Radio Access Network (O-RAN) has emerged as a critical architecture for future cellular infrastructure. As bandwidth demands continue to surge, it is common that O-RAN small cells operate on the same frequency band (i.e., frequency reuse factor of 1) to maximize spectral efficiency \cite{oran2023vertical}. However, the small-cell architecture exacerbates inter-cell interference, posing a significant challenge that requires careful interference management to ensure reliable performance. 
While extensive research exists on inter-cell interference management (ICIM), the innovative architectural paradigm of O-RAN unlocks new opportunities for deploying online learning approaches in real-world RAN systems by leveraging the xApps hosted within the Near-Real-Time (Near-RT) RAN Intelligent Controller (RIC). These approaches can rapidly adapt to interference conditions in the network to maximize the user-defined objectives while respecting the quality of service (QoS) demands of users.

Diffusion models have demonstrated remarkable capabilities in content generation across diverse domains, such as producing high-quality images, videos, and text by iteratively refining noisy data into structured output. This generative prowess stems from their ability to model complex, high-dimensional distributions, making them a powerful tool beyond traditional applications. In the context of online interference management in O-RAN, diffusion models offer a compelling fit due to their capability to capture the stochastic nature of inter-cell interference and adaptively refine resource allocation policies. 
% diffusion models provide a compelling solution due to their ability to capture the stochastic nature of inter-cell interference and adaptively refine resource allocation policies.
% Their iterative denoising process aligns with the dynamic, uncertain environment of small-cell O-RAN, where frequency reuse intensifies interference, necessitating a robust policy for interference-aware resource allocation. 
Despite their potential, diffusion models remain underexplored in O-RAN applications, as existing interference management strategies largely rely on conventional rule-based or learning-based approaches. This underutilization highlights an opportunity to harness the unique strengths of diffusion models to generate robust policies for the online optimization of O-RAN systems.

In this paper, we study the ICIM problem in an O-RAN system where adjacent small cells operate on the same frequency band.\footnote{
We note that the frequency reuse factor of 1 is commonplace in real-world cellular networks as the operators push the boundary of spectrum utilization, especially in densely populated areas (see \cite{k2024unveiling, hassan2022vivisecting, 10.1145/3452296.3472923}).}
We propose \pname, a diffusion-based reinforcement learning (RL) framework that generates policies to guide the resource allocation at O-RAN distributed units (DUs).
\pname operates as an xApp within the Near-RT RIC, periodically querying DUs via the E2 interface to collect key performance metric (KPM) and media access control (MAC) data. Based on the KPM and MAC data observations, it dynamically refines its policies to guide the resource allocation at individual DUs, aiming to maximize a user-defined reward function while accounting for inter-cell interference, time-varying channel conditions, and dynamic user QoS demands.

One challenge in the design of \pname lies in the time-scale discrepancy between RIC and DU operations.
The Near-RT RIC updates its resource management policies based on the KPM and MAC data from multiple DUs, operating at a near-real-time scale of 10~ms to 1~s. In contrast, DUs operate at a real-time scale of 1 ms, performing subframe-by-subframe resource allocation based on the policies provided by the Near-RT RIC.
In addition to this time-scale mismatch, individual DUs cannot cooperate in real time for joint interference management due to inter-DU communication latency.
Instead, cooperation and coordination among DUs need to be achieved through the policies generated by the Near-RT RIC.

To address this challenge, we introduce a new metric called the \textit{preference value} for each resource block (RB) at every DU. This metric mathematically represents the policy generated by the Near-RT RIC, enabling interference-aware resource allocation at individual DUs.  
The numerical policy representation serves as a crucial bridge between near-real-time and real-time operations for efficient ICIM. For the Near-RT RIC, it simplifies the output format of the diffusion model, accelerating the convergence of its training process. This is particularly important for online learning, as the diffusion model must be continuously trained based on KPM and MAC data observations to adapt to network dynamics, such as interference conditions, time-varying channels, and fluctuating user demands.  
For individual DUs, this numerical policy representation acts as a set of scheduling-priority weights for resource block scheduling. It can be seamlessly integrated with existing DU scheduling algorithms, such as proportional fairness (PF).

% Specifically, we adopt a two-step resource allocation strategy that leverages \textit{preference values} for the joint control of multiple cells.
% In the first step, online learning agents within the Near-RT RIC continuously monitors and updates the \textit{preference values} for individual O-DUs based on network conditions and user demands.
% The second step involves real-time scheduling within individual O-DUs, where the system first checks the availability of resource block groups and then makes resource allocation decisions based on the received \textit{ b} to maximize a user-defined reward function.

Another challenge within the design of \pname is the integration of the diffusion model with reinforcement learning.
Unlike image and video generation tasks, where diffusion models are trained offline on large datasets to generate high-quality outputs without requiring continuous adaptation, online policy generation in O-RAN requires that diffusion models continuously adapt to dynamic network conditions. 
Moreover, unlike static image/video datasets, the input distribution of diffusion models in O-RAN evolves in real-time based on user demands, interference levels, and network states. 
This requires continuous training and updates of diffusion models using streaming KPM and MAC data from DUs.

To address this challenge, we propose an efficient architecture that integrates a diffusion model with RL for adaptive policy generation. Specifically, we use a conditional denoising diffusion probabilistic model (DDPM) to generate policies and employ a critic with double Q-learning networks to evaluate them.  
One difficulty lies in effectively aligning the diffusion model's generative process with the policy optimization objectives of Q-learning. To overcome this, we introduce a joint training approach that combines the Q-learning loss with the denoising loss. By incorporating both losses, the diffusion model learns to generate policies that not only capture the underlying data distribution but are also optimized for improved performance based on the critic's feedback.  
Moreover, this architecture leverages the strengths of both policy-based and value-based models, reducing the number of denoising steps required by the diffusion model. This makes it well-suited for Near-RT operations.

We have integrated \pname into the OpenAirInterface (OAI) software suite and evaluated its performance on a 5G testbed consisting of three cells and a set of smartphones.
Ablation studies show that the diffusion model plays a key role in generating policies to improve network performance in the face of inter-cell interference. 
% Particularly, we compare its performance against the state-of-the-art models, including deep Q-learning and actor-critic models, in two O-RAN small-cell scenarios. 
% Diffusion models enhance policy robustness by leveraging structured generative processes, thereby addressing the high variance and instability of traditional RL methods. 
%While RL approaches often exhibit slow convergence in dynamic environments, diffusion-based optimization enables rapid adaptation to changing network conditions and efficiently estimates preference values. %Peihao
%\hz{more insights needed here... Done}
Extensive experimental results demonstrate that \pname is robust in Near-RT policy generation and outperforms the state-of-the-art (SOTA) RL approaches, showcasing the potential of diffusion models for online optimization in O-RAN.

This work advances the state-of-the-art as follows.
\begin{itemize}
\item 
We propose a new metric called \textit{preference value} as the policy representation for the Near-RT RIC to manage inter-cell interference.  
This compact representation not only simplifies the training process of the diffusion model but also seamlessly integrates with existing DU scheduling algorithms.

\item 
We propose a framework for integrating diffusion model with RL. It appears to be robust and efficient in maximizing user-defined reward functions in the presence of inter-cell interference. 

\item 
Extensive experimental results demonstrate that \pname outperforms the SOTA ICIM approaches, including deep Q-learning and actor-critic RL.

\end{itemize}

\section{Related Work}

Interference is a fundamental problem in wireless networks. 
Even within cellular networks, the literature already has a large body of work on interference management, 
% ranging from theoretical analysis \cite{6815893, wei2019multi, lopez2012expanded} and protocol design \cite{,,,,,} to experimental validation \cite{,,,,,}. 
% To mitigate severe ICI, interference management strategies have been explored in the different domains, 
encompassing power-domain methods \cite{lopez2012expanded, lopez2013power}, time-domain techniques \cite{altay2021design, ayala2019online}, frequency-domain approaches \cite{yan2021self}, spatial-domain solutions \cite{chaudhari2022machine}, and code-domain strategies \cite{li2015scma}.
Since it is impossible to review all existing work, our survey focuses on ICIM in 5G and O-RAN. 

% Inter-cell Interference (ICI)\cite{yu2014combating, zou2024trident, ranjbar2022cell} remains a significant challenge in 5G networks\cite{hassan2022vivisecting}, impacting user experience and overall network performance. 
% 4g, 5g

\textbf{ICIM in 5G and O-RAN.}
A number of ICIM techniques have been proposed for 4G/5G cellular networks, such as enhanced inter-cell interference coordination (eICIC) \cite{deb2013algorithms}, coordinated multi-point (CoMP) \cite{lee2012coordinated}, and dynamic power control \cite{5949138}. In recent years, the use of massive MIMO, beamforming, and carrier aggregation technologies has gained increasing attention for interference suppression \cite{li2023ca++, ye2024dissecting,9828487}.
Meanwhile, both data-driven optimization approaches \cite{liu2015joint, sohaib2024drl} and learning-based techniques \cite{elsayed2020machine, yan2021self, 9791337} have advanced in increasingly sophisticated forms to enhance the management of inter-cell interference, leading to significant improvements in spectrum utilization for 5G networks \cite{ye2024dissecting, ngo2024ai, akgun2024interference}.

The new architecture of O-RAN presents opportunities to deploy intelligent solutions for ICIM in realistic 5G networks, driving the rapid development of DNN models \cite{anand2024machine, 10211189, 10623003, akgun2024interference}, RL solutions \cite{simsek2012dynamic, 10622781, xiao2019reinforcement, elsayed2020transfer, hu2022inter}, and graph-based models \cite{gu2024graph} for interference prediction and management.
The flexibility of O-RAN further enables inter-cell coordination via the RIC for interference mitigation \cite{5351727, 10330565, baldesi2022charm, schiavo2024cloudric, ko2024edgeric}, enhancing 5G network resilience, robustness, and scalability.

% In O-RAN, ICIM benefits from the open and intelligent architecture, enabling non-real time, near-real time, and real-time optimization. 
% AI-driven solutions \cite{li2024architecture}, such as machine learning \cite{anand2024machine, 10211189, 10623003, akgun2024interference}, reinforcement learning \cite{simsek2012dynamic, 10622781, xiao2019reinforcement, elsayed2020transfer, hu2022inter} and graph-based models \cite{gu2024graph}, are increasingly used to predict interference patterns \cite{10773839, 10773685} and optimize resource allocation \cite{tanaka2024ran} dynamically. 

% The flexibility of O-RAN also allows for cross-DU coordination \cite{5351727, 10330565, baldesi2022charm, schiavo2024cloudric, ko2024edgeric} via RIC, improving interference mitigation while maintaining scalability and interoperability.

% \hz{the above two paragraphs need to be reorganized and better structured to reflect the current knowledge in the literature. Done}

\begin{table}
\scriptsize
\centering
\caption{5G network ICI approaches in the literature.}
\resizebox{\linewidth}{!}{
\begin{tabular}{|l|l|l|l|l|l|}
\hline
%\textbf{Type}&
\textbf{Reference} & \textbf{Objective} & \textbf{Key Idea} & \textbf{\!\!\!\!OTA?\!\!\!\!} & \textbf{\!\!\!UE\!\!\!} & \textbf{\!\!\!\!RIC?\!\!\!\!}  \\ \hline

\makecell[l]{eICIC\cite{deb2013algorithms} \\ CoMP\cite{lee2012coordinated}\\ lte-a\cite{liu2015joint}} & Spectrum Efficiency  & \makecell[l]{Almost Blank Subframe\\(ABS)} & \ding{51} & \ding{55} & \ding{55}\\ \hline
% from a real network deployment 
%Maximize Deliverable Streams
CoaCa\cite{yu2014combating} & Maximize Streams  & Antenna Beamforming & \ding{51} & \cite{MARP_ue}  & \ding{55}\\ \hline

CSRS\cite{wu2022demand}&  feICIC & FAP clustering & \ding{55} & \ding{55}  & \ding{51}  \\ \hline

2-Layer IC\cite{ge20222}& Improve Link Capacity & RCGCA &  \ding{55} & \ding{55} & \ding{51} \\ \hline

% Qualcomm\cite{Qualcomm_ue, Qualcomm_cell}
IAIS\cite{akgun2024interference} &  Enhance reliability & \makecell[l]{Interference prediction} & \ding{51} &  \cite{Qualcomm_ue}  & \ding{51} \\  \hline

\makecell[l]{ ChARM\cite{baldesi2022charm} \\ IM-rApp\cite{ngo2024ai} \\ 
}&  Improve Throughput & \makecell[l]{Machine Learning \\ (Prediction)} & \ding{51} &  N/A  & \ding{51} \\  \hline

MLMCOS\cite{anand2024machine}&  Maintain High QoE & ML Classification & \ding{55}    & \cite{srsRANue}  & \ding{51} \\  \hline

\makecell[l]{DRL-IM\cite{sohaib2024drl} \\ mmLBRA\cite{10293795}\\  Dynamic-IC\cite{simsek2012dynamic}
}& Increase Sum-Rate & Reinforcement Learning & \ding{55} & \ding{55} & \ding{51}\\   \hline

%  X60 5G Modem
\textbf{\pname (Ours)}& Improve System QoS & Diffusion Policy & \ding{51} & \ding{51} & \ding{51}\\ \hline
% generate interference map to O-DU
\end{tabular}
}
\label{tab:relwork}
\vspace{-0.15in}
\end{table}

Although the literature contains a large body of work on ICIM, little progress has been made in the design of online learning solutions and their validation in realistic networks.  
\pname fills this gap.  
Table~\ref{tab:relwork} highlights the most relevant work and positions \pname uniquely in the literature.  
Specifically, \pname differs from prior work in that it is the first to use diffusion policies for ICIM and has been validated on realistic 5G testbeds.

\textbf{xApps for Resource Allocation in O-RAN.}
% Resource allocation is The literature contains a substantial body of work on MAC-layer resource allocation for cellular networks (see, e.g., \cite{5351727, 9355403}).
% \hz{references needed}. 
% Recently, research efforts have shifted toward O-RAN systems \cite{chen2023om,li2024architecture, ranjbar2022cell, bai2024distributed, 10620908, abdalla2022toward}, particularly through the development of xApps in the Near-RT RIC \cite{schiavo2024cloudric,ko2024edgeric}.  
%
Many xApps have been developed to optimize various network performance metrics---such as spectral efficiency, latency, fairness, and reliability---by dynamically enforcing resource allocation policies at DUs (see, e.g., \cite{5351727, 9355403}). Among these approaches, online learning \cite{aslan2024fair}, particularly RL \cite{10293795, simsek2012dynamic}, has emerged as an appealing technique for xApps due to its ability to adaptively refine resource allocation policies in response to time-varying network conditions.  
Actually, a considerable volume of work has explored RL for network management and optimization in O-RAN, including network slicing \cite{hu2022inter, filali2022dynamic}, flow scheduling \cite{bai2024distributed}, MCS selection \cite{an2024dragon,jonnavithula2024mimo}.
However, despite these prior efforts, limited progress has been made in developing RL-based xApps for ICIM in O-RAN.  
\pname fills this gap by introducing a novel RL-based xApp for ICIM.

\textbf{Diffusion Policy for Reinforcement Learning.}
Diffusion models, originally developed for image and video synthesis \cite{ho2020denoising}, have recently gained attention in RL for their ability to model complex, high-dimensional policy distributions \cite{zhu2023diffusion, chi2023diffusion, wang2022diffusion}.
Unlike traditional RL approaches that rely on deterministic or stochastic policy networks, diffusion models leverage a generative process that iteratively refines noisy inputs into structured outputs, enabling more expressive and flexible policy representations. In an RL framework, diffusion models can be used to learn a distribution over optimal actions, enhancing agents' adaptation in dynamic environments. 
For instance, pioneering work has integrated diffusion models with RL to model complex distributions of possible actions \cite{rigter2023world, mazoure2023value, yang2023policy}, make multi-step predictions about the next state \cite{janner2022planning, he2023diffusion}, and expand the database \cite{lu2023Synthetic}. 
To the best of our knowledge, diffusion models have not been explored for online resource optimization in O-RAN. 

\section{Problem Statement}\label{sec:problem}

% why multi DU rather than multi RU.

% - multiple DUs from multiple carriers. 
% - for multi-RU solution, the data capacity of fronthaul is huge (due to the transfer of frequency signal).

\begin{figure}
    \centering
    \includegraphics[trim= 10 10 10 10, clip, width=\linewidth]{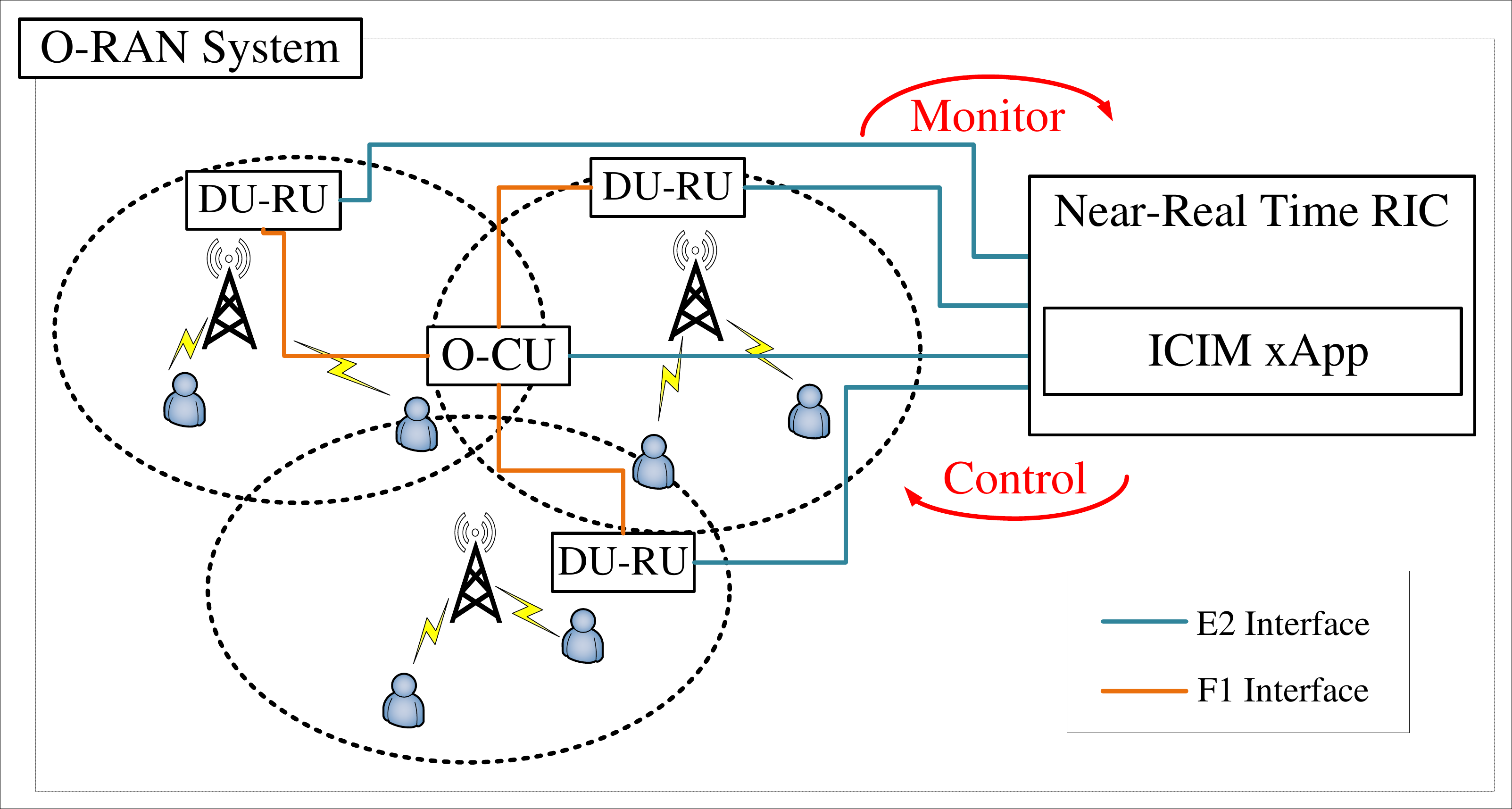}
    \caption{Inter-cell interference among O-RAN small cells.}
    % \hz{this figure is too busy. simplify it to emphasize the key components. Done.}
    \label{fig:xdiff_introduce}
    % \vspace{-0.2in}
\end{figure}

Consider an O-RAN system as shown in 
Fig.~\ref{fig:xdiff_introduce}, where the small cells operate on the same frequency band for both uplink and downlink transmissions. 
Each small cell is equipped with one RU and one DU. 
A central RIC is connected to all DU devices for remote control and management. 
Due to the physical distance between RIC and DU, the control and management of DUs can only be executed at the Near-RT level. 
The primary role of RIC in this architecture is to generate Near-RT policies that guide resource allocation operations in individual DUs, optimizing the user-defined reward function. 
\mycut{
This approach enables dynamic and adaptive resource management while maintaining low-latency control, which is essential for optimizing network performance in multi-cell O-RAN deployments.
}
For this network setting, we have the following two notes.

% \subsection{Background}
\textbf{Frequency Reuse.}
Given the limited availability of spectrum, frequency reuse is essential for accommodating the growing demand for wireless services. By reusing the same frequency in adjacent cells, operators can deploy a higher density of base stations, such as small cells, to improve coverage and capacity without requiring additional spectrum resources.
In fact, real 5G networks commonly reuse the same frequency bands for neighboring cells---a practice known as frequency reuse or universal frequency reuse (Reuse-1). 
Ericsson's study shows that Reuse-1 will play an important role in meeting the growing capacity demands in dense 5G networks and has the potential to double the network capacity compared to Reuse-2
\cite{ericsson_frequency_reuse}.
The aggressive frequency reuse calls for interference management techniques that enable adjacent cells to operate on the same frequency band.

% Unlike earlier cellular generations that often used different frequency bands in adjacent cells to minimize interference (e.g., Frequency Reuse-3 or Reuse-7 patterns), 5G relies on advanced interference management techniques to enable adjacent cells to operate on the same frequency band. 

% On one hand, the risk of interference between neighboring links is much higher in a Reuse 1 network when compared to a Reuse 2 network, but, on the other hand, the channel bandwidth allocated to each link is doubled. Therefore, the trade-off between interference and bandwidth becomes of high interest. Traditionally, backhaul links operate in the bandwidth-limited regime where the signal-to-interference-and-noise ratio (SINR) is very high. However, in the bandwidth-limited regime, the link capacity scales linearly with bandwidth and only logarithmically with SINR. Interference will reduce SINR, but the key point is that it is much more beneficial to take a slight penalty in SINR in order to be able to enjoy a wider bandwidth when the system is bandwidth-limited.

% We note that the frequency reuse factor of 1 is commonplace in real-world cellular networks as the operators push the boundary of spectrum utilization, especially in dense population areas (see \cite{k2024unveiling, hassan2022vivisecting}). 

\textbf{Multi-DU Multi-Cell Architecture:}
The flexibility of O-RAN allows for the support of multiple cells (i.e., multiple RUs) using either a single DU or multiple DUs.
In the former case, signals from all RUs are jointly processed at a single DU, making it well-suited for CoMP and massive MIMO transmissions. However, this approach places a heavy computational burden on the DU and requires high-capacity front-haul links, posing challenges for practical deployment.
In the latter case, PHY-layer signal processing is performed independently at individual DUs, with coordination managed at the Near-RT RIC. While this architecture does not fully optimize network capacity, it offers great flexibility in network operation and maintenance.
In this work, we consider the multi-DU architecture for our design.

\vspace{-0.15in}
\subsection{Experimental Observations}

\begin{figure*}[!t]
    \centering
    \hspace{0\textwidth} 
    \begin{subfigure}[b]{0.3\textwidth}
        \centering
        \includegraphics[trim=0 0 0 0, clip, width=\linewidth]{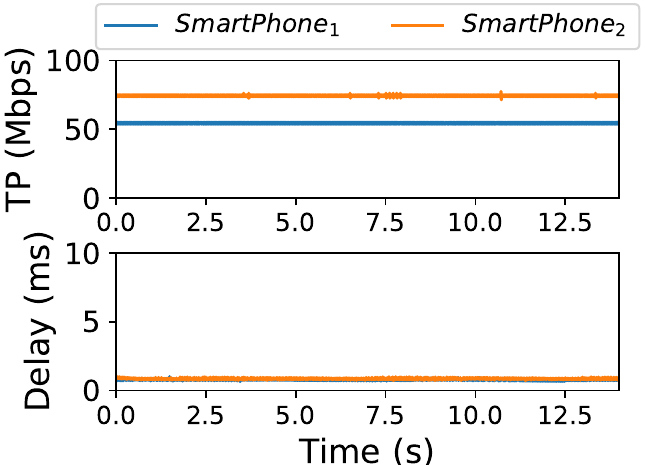}
        \caption{No inter-cell interference.}
        \label{fig:nointerference}
    \end{subfigure}
    \hspace{0\textwidth}  
    \begin{subfigure}[b]{0.3\textwidth}
        \centering
        \includegraphics[width=\linewidth]{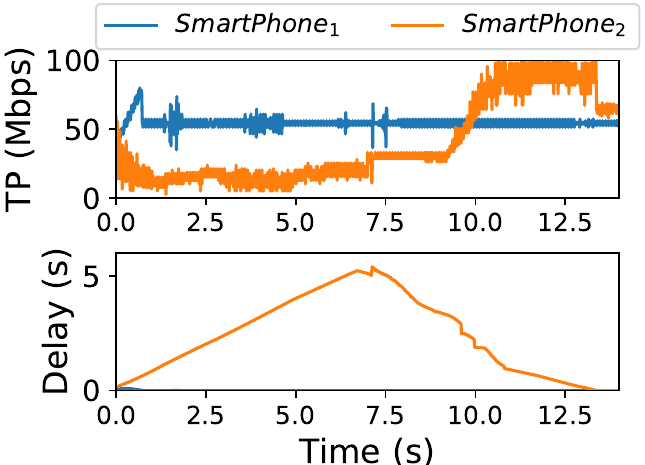}
        \caption{W/o policy from RIC.}
        \label{fig:highlatency}
    \end{subfigure}
    \hspace{0\textwidth}  
    \begin{subfigure}[b]{0.3\textwidth}
        \centering
        \includegraphics[width=\linewidth]{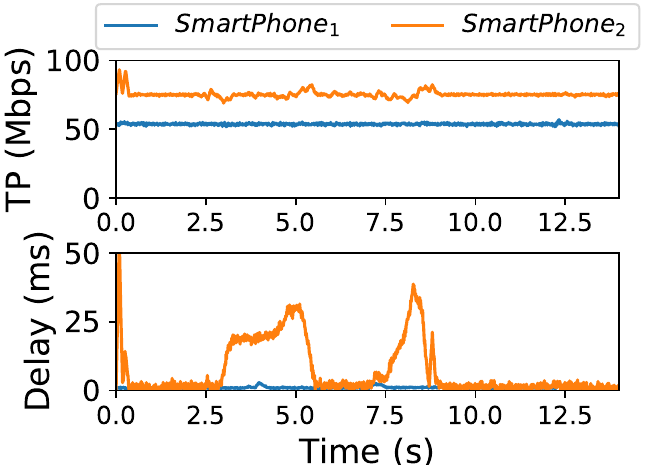}
        \caption{W/ a simple policy from RIC.}
        \label{fig:simpleric}
    \end{subfigure}
        \caption{Impacts of inter-cell interference on user throughput (TP) and queue delay.}
        % latency (queue delay).
        \label{fig:effectICI}
        % \vspace{-0.15in}
\end{figure*}

To understand the impacts of inter-cell interference, we conduct experiments on an OAI testbed that comprises two cells and two commercial smartphones (see details in \S\ref{sec:implementation}).
The core network generates persistent downlink data traffic at 50~Mbps for smartphone 1 and at 60~Mbps for smartphone 2 using \verb|iperf|. 
The real throughput (TP) and queue delay of both smartphones were measured at their respective DUs. 
Fig.~\ref{fig:effectICI} presents our observations in three cases. 
\begin{itemize}
\item 
\textbf{Case I:} We intentionally avoid inter-cell interference by assigning the two cells to different frequency bands: one on n78 and the other on n41.  
Fig.~\ref{fig:nointerference} presents our measurement results.  
Evidently, the two smartphones maintain stable and reliable link connections with an acceptable delay of 1~ms.  

\item 
\textbf{Case II:} We assign both cells to the same frequency band, i.e., n78.  
Fig.~\ref{fig:highlatency} shows our measurement results.  
It can be seen that smartphone 2 experiences an unstable link connection with significant queue delays (up to 5~s).  
This connection instability is caused by interference from the other cell.  

\item 
\textbf{Case III:} 
In contrast to Case II, we use the RIC to deploy a simple resource allocation policy for ICIM: cell 1 uses 0--50~RBs while cell 2 uses 51--106~RBs.  
Fig.~\ref{fig:simpleric} presents our measurement results in this case.  
It shows that both smartphones maintain stable connections with acceptable delay. 
\end{itemize}
These experimental results demonstrate the destructiveness of inter-cell interference and showcase the effectiveness of the RIC policy in ICIM for meeting users' throughput and delay demands.

\subsection{Reward-Based Formulation}

The objective of this work is to develop an xApp for Near-RT RIC that can generate an efficient policy to guide the resource allocation of DUs in the presence of inter-cell interference. 
Denote $\mathcal{K}$ as the set of small cells, with $RU_k$ and $DU_k$ denoting its radio and distributed units. 
Denote $\mathcal{U}_k$ as the set of UEs served by $DU_k$, with $\mathcal{U} = \bigcup_{k \in \mathcal{K}} \mathcal{U}_k$ denoting the set of all UEs in the network. 
While DU performs resource allocation subframe-by-subframe at the time scale of 1~ms, the policy from the RIC updates at a slower pace, with the time scale from 10~ms to 1~s. 
\mycut{We assume that the time slot is the time granularity of policy updates from the Near-RT RIC. 
}
%\hz{this sentence is not clear.}

For UE $i \in \mathcal{U}$, denote $P_i$ as its throughput demand. 
Denote $\rho_i(t)$ as its achieved throughput in time slot $t$, where $t = 0, 1, 2, \dots$.
If $\rho_i(t) \ge P_i$, then the UE's throughput demand is met and the regret is zero; otherwise, we define the regret as its normalized throughput deficit, 
i.e., \( \frac{P_i-\rho_i(t)}{P_i}\).
Combining these two cases, we model the regret of UE $i$ in time slot $t$ as follows:
$\max\big(\frac{P_i-\rho_i(t)}{P_i}, 0\big)$.

Similarly, for UE $i \in \mathcal{U}$, denote $D_i$ as its delay demand. 
Denote $\tau_i(t)$ as the achieved average delay of UE $i$'s data packets in time slot $t$.
If $\tau_i(t) \le D_i$, then the UE's delay demand is met and its delay regret is zero; otherwise, we define the delay regret as its normalized delay deficit, 
i.e., \( \frac{\tau_i(t) - D_i}{D_i}\).
Combining these two cases, we model the delay regret of UE $i$ in time slot $t$ as follows:
$\max\big(\frac{\tau_i(t) - D_i}{D_i}, 0\big)$.

% We formulate this optimization problem as a maximization reward problem:
% Let \(T\) denote the time duration of one query round and \( t \) denote xApp's query iteration index. 
% For UE \( i \in \)  gNB $k$ , let \( \rho_i(t) \) denote its achieved throughput, $\mathrm{[P]}$ denote its throughput demand, \( \tau_i(t) \) denote its experienced time delay, $\mathrm{[D]}$ denote its delay requirement, all within query round \( t \).
% % All these KPMs are computed in RAN and reported to xApp in response to its queries. 

% We now model the reward of UE $i$ $\in$ gNB $k$ and time slot $t$, which includes two components: throughput and delay.
% For throughput, if its achieved throughput is greater than its demand, the reward is zero; otherwise, the reward is defined as the normalized throughput deficit (i.e., $\frac{\rho_i(t) - P_k}{P_k}$). 
% Combining these two cases, its throughput reward is modeled as: 
% $\min\big(\frac{\rho_i(t) - P_k}{P_k}, 0\big)$. 
% Similarly, the delay is modeled as:
% $\min\big(\frac{D_k - \tau_i(t)}{D_k}, 0\big)$.

Following the convention, we use reward instead of regret as our optimization objective function. 
To do so, we define the reward value as the inverse of the regret value. 
Specifically, let $r_{k}^{\mathrm{[p]}}(t)$ denote the throughput reward of small cell $k$ in time slot $t$,
and
$r_{k}^{\mathrm{[d]}}(t)$ denote its delay reward.
Then, we have:
\begin{align}
& r_{k}^{\mathrm{[p]}}(t) = \sum_{i \in \mathcal{U}_k} \min\big(\frac{\rho_i(t) - P_i}{P_i}, 0\big),  \\
& r_{k}^{\mathrm{[d]}}(t) = \sum_{i \in \mathcal{U}_k} \min\big(\frac{D_i - \tau_i(t)}{D_i}, 0\big), 
\end{align}
where both 
\(r_{k}^{\mathrm{[p]}}(t)\) 
and
\(r_{k}^{\mathrm{[d]}}(t)\) 
are non-positive values. 

To model the varying QoS demands of different UEs, we introduce a non-negative vector to denote the weights of their throughput and delay rewards. 
Specifically, let $\lambda_{k}^{\mathrm{[p]}}$, $\lambda_{k}^{\mathrm{[d]}}$ denote
the throughput and delay weights in small cell $k$. 
Network operators can use these weights to adjust the priorities of throughput and delay during online optimization.
Incorporating the weights, the QoS metric, representing the total reward of all UEs in all cells, can be written as:
\begin{equation}
\!\!r(t) 
\!=\! 
\sum_{k \in \mathcal{K}} 
\left(
\lambda_{k}^{\mathrm{[p]}} r_{k}^{\mathrm{[p]}}(t)
+
\lambda_{k}^{\mathrm{[d]}} r_{k}^{\mathrm{[d]}}(t)
\right) 
\,.
\label{eq:reward_all}
\end{equation}

Based on the above reward functions, we formulate this problem as a Markov decision process (MDP).
The objective is to find the optimal policy $\pi^*$ that maximizes the expected cumulative reward (discounted sum of rewards), i.e.,
\begin{equation}
\pi^* = \arg\max_{\pi} \mathbb{E} \left[ \sum_{t=0}^\infty \gamma^t r(t) \right].
\label{eq:opt}
\end{equation}

In this optimization problem, DUs are the environment; Near-RT RIC is the AI agent that makes policy decisions. 
UE QoS demands $P_i$ and $D_i$, $i \in \mathcal{U}$, are given values that may change over time. UE QoS metrics $\rho_i$ and $\tau_i$ are observable variables. 
Throughput and delay weights, $\lambda_{k}^{\mathrm{[p]}}$ and $\lambda_{k}^{\mathrm{[d]}}$, are given constants for all $k \in \mathcal{K}$.

\section{\pname: Design}

\label{subsec:overview}

\FloatBarrier
\begin{figure*}[!t]
    \centering    \includegraphics[width=\linewidth, trim= 0 0 0 0, clip]{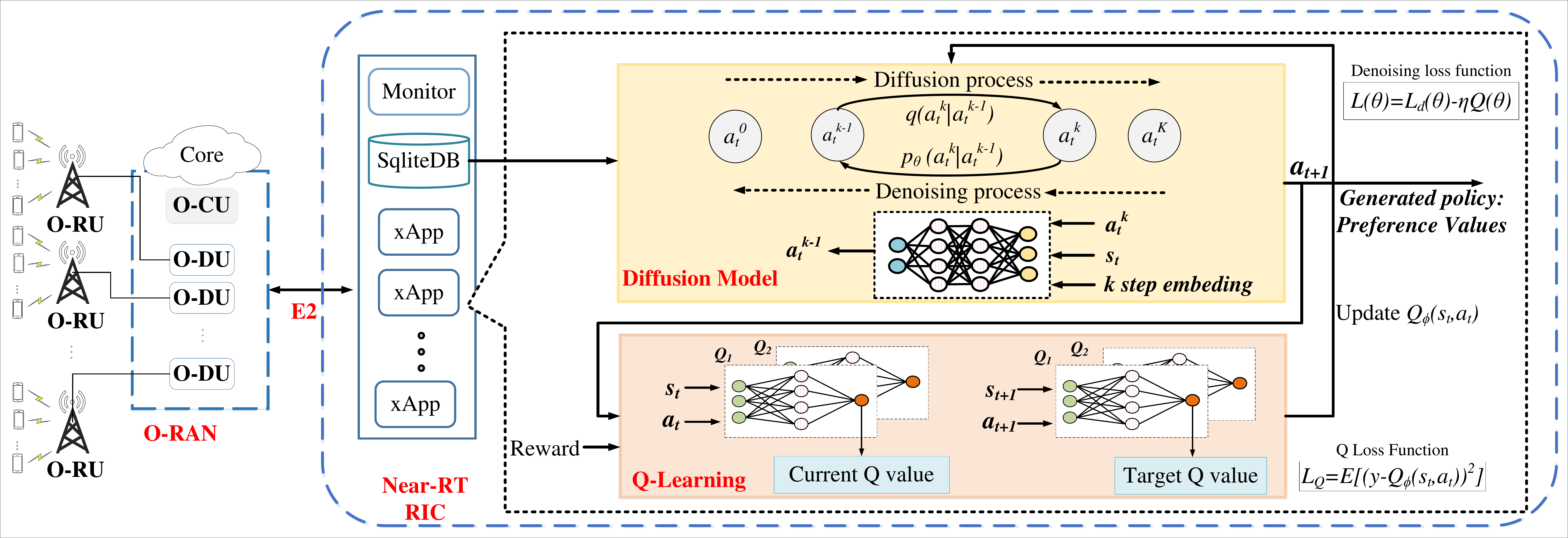}
    \caption{Architecture of \pname.}
    \label{fig:framework}
\end{figure*}

We propose \pname to address the optimization problem in Eq.~\eqref{eq:opt}. 
Fig.~\ref{fig:framework} shows the key components of \pname in the architecture of O-RAN. 
\pname is an xApp residing within the Near-RT RIC, generating policies to guide the resource allocation at individual DUs. 
It should be stressed that the policy is generated by \pname in Near-RT fashion (in the time scale of 10 ms to 1 s), while the resource is allocated at the DUs in real-time fashion (every one millisecond). 
Model input, output, and structure are three key components of \pname, which we describe as follows.

% \usepackage{multirow}
\begin{table}[!t]
% \hz{I copied this table from previous paper. make sure it is right.}
\caption{The list of KPM and MAC data that \pname obtains for each UE.}
\resizebox{\linewidth}{!}{
\begin{tabular}{|c|l|l|l|}
\hline
\multicolumn{1}{|l|}{} & Data & \!\!UL or DL?\!\!\! & Explanation \\ \hline
\!\!\!\multirow{3}{*}{\begin{tabular}[c]{@{}c@{}}KPM \\ data\end{tabular}}\!\!\! & Per-UE TP\!\!\! & DL & Average data rate achieved by UE  \\ \cline{2-4} 
% in 10 ms. 
 & Per-UE delay & DL & Delay of SDU after being requested. \\ \cline{2-4} 
 %delay between MAC requesting data and RLC delivering it for transmission.
 & Per-UE PRBs & DL & \# of PRBs assigned to each UE in 10 ms.\\ \hline
\!\!\!\multirow{7}{*}{\begin{tabular}[c]{@{}c@{}}MAC \\ data\end{tabular}}\!\!\! & PUSCH SNR & UL & Quality of signal transmitted by UE. \\ \cline{2-4} 
 & PHR & UL &  Max Tx power - current usage power. \\ \cline{2-4} 
 & MCS & DL\&UL &  Modulation index and coding rate for data. \\ \cline{2-4} 
 & BLER & DL\&UL & Percentage of blocks received with errors. \\ \cline{2-4} 
 & Current TBs & DL&  \# of Transport Blocks being Tx-ed.\!\!\!\! \\ \cline{2-4}  % in a given time
 & Scheduled RBs & DL& \# of PRBs scheduled for transmission. \\ \hline  %\cline{2-4} 
% &  &  &  \\ \cline{2-4} 
% &  &  &  \\ \hline
\end{tabular}
}
\label{tab:kpmmacdata}
\vspace{-0.2in}
\end{table}

\textbf{Model Input.}
\pname obtains the RAN state information from the DUs via the E2 interface and uses the information as the input to infer the policy for the resource allocation of individual DUs. 
The RAN state information includes both KPM and MAC data at each DU. 
Table~\ref{tab:kpmmacdata} lists the KPM and MAC data that are used as input to \pname. 
Denote $B$ as the number of (KPM and MAC data) samples that \pname obtains per second from one DU. 
We observed that $B$ varies from 100 to 1000 samples per second.

\textbf{Model Output.}
The model output is a resource allocation policy for individual DUs, which plays a critical role in the management of inter-cell interference. 
The time-scale discrepancy between RIC and DU operations makes it infeasible for RIC to directly manage resource allocation for individual DUs, necessitating an efficient policy representation to bridge the near-real-time and real-time operations.  
The policy representation should capture both short-term and long-term network dynamics, including interference levels, channel conditions, user mobility, and user demands, to guide real-time resource allocation at DUs. Additionally, it should be lightweight to minimize communication overhead on the E2 interface.  
To address these challenges, we propose an elegant mathematical formulation as the policy representation, which will be explained in \S\ref{subsec:policy_rep}.

\textbf{Policy Agent Structure.}
Even with a well-defined policy representation, the policy agent design is a challenging task as it must generate a policy in a near-real-time manner. The policy agent must rapidly adapt to network dynamics while meeting stringent latency requirements. Additionally, it must ensure fast policy convergence, allowing the learning algorithm to quickly adjust to time-varying network conditions without prolonged training periods. Furthermore, due to the physical separation between RIC and DU, the policy agent in RIC operates with partial observability and incomplete network information, necessitating an efficient learning architecture for policy generation.
To address the above challenges, we propose a diffusion model for policy generation within an RL framework, as shown in Fig.~\ref{fig:framework}.
Details will be presented in \S\ref{subsec:diffusion}.

% We designed an algorithm for interaction and joint decision-making between the Near-RT RIC and the DU.
% Fig. \ref{fig:framework} presents the system architecture of \pname in an O-RAN. \pname is designed as an xApp running in the Near-RT RIC, communicating with gNB (O-CU, O-DU, and O-RU) via the well-established E2 interface
% and generating interference map for O-DU. It obtained the KPM and MAC data from each O-DU via the E2 interface for each individual UE in each iteration. 
% These data is used as the input of \pname to update its interference map with the aim of maximizing the reward defined in Equation (\ref{eq:reward_all}).
% Fig. \ref{fig:PFscheduler} presents the system architecture of O-DU.

\subsection{Policy Representation}
\label{subsec:policy_rep}

% \noindent
% \textbf{Basic Idea:}
Each DU is responsible for the subframe-by-subframe (real-time) resource allocation of its own cell, in the presence of inter-cell interference. 
Due to the inter-DU communication latency, joint resource allocation cannot be achieved in a real-time fashion. 
Therefore, the DUs need to be coordinated by the Near-RT RIC for interference-aware resource allocation.

% \subsection{Preference Values from Near-RT RIC}

\textbf{Hard Policy.}
A natural approach for the Near-RT RIC to coordinate resource allocation is by controlling the use of resource blocks at each DU.  
Specifically, the policy agent generates a \textit{preference value} for each UE in each cell on each RB,
i.e., 
\begin{equation}
p(DU_k, UE_i, RGB_j) =
\begin{cases}
1, & \!\!\!\!\text{\scriptsize For $DU_k$, allocate $RB_j$ to $UE_i$,} \\
0, & \!\!\!\!\text{\scriptsize No preference, decide it by $DU_k$,} \\
-1, & \!\!\!\!\text{\scriptsize For $DU_k$, do not allocate $RB_j$ to $UE_i$,}\!\!\!\!\!\!\!\!  
\end{cases} 
\label{eq:map}
\end{equation}
where $i \in \mathcal{U}_k$, $k \in \mathcal{K}$, and $j \in \mathcal{J}$.
For instance, $\mathcal{J} = \{1, 2, \dots, 106\}$ for 5G NR with 40~MHz bandwidth. 
Through learning the interference patterns in the network, the policy agent intends to use ``1'' and ``-1'' as the recommendation values to avoid strong inter-cell interference on the same RB. 
If there is no strong inter-cell interference on an RB, the policy agent does not make a recommendation for this RB, and the DU can allocate this RB using its own scheduling algorithm. 
The Near-RT RIC sends the generated recommendation values to each DU on the E2 interface.
Upon receiving the recommendation values, $DU_k$ follows the recommendation for its resource allocation.

\textbf{Soft Policy.}
While this policy representation has a simple format, it does not perform well in our experimental tests. 
We observed that DUs frequently struggled to find sufficient RBs for UE scheduling, resulting in a high delay of UE communications. 
This may be attributed to its \textit{hard} recommendation reflected by its discrete policy representation values, which explicitly excludes a subset of RBs for the UEs and thus limits the scheduling flexibility of DUs.

Based on the experimental observations, we relax the policy representation values from discrete numbers to continuous numbers within the range from -1 to 1,
i.e.,
\begin{equation}
p(DU_k, UE_i, RB_j) 
\in [-1, ~1].
\label{eq:map2}
\end{equation}

A large value means that an RB has a high preference for the UE, while a small value means the RB has a low preference for the UE.
We thus call the policy representation as \textit{preference values}.
This \textit{soft} policy allows for flexible real-time resource allocation at the DUs while respecting the recommendation from the policy agent.

\begin{figure}
    \centering
    \includegraphics[trim=0 0 0 0, clip, width=0.7\linewidth]{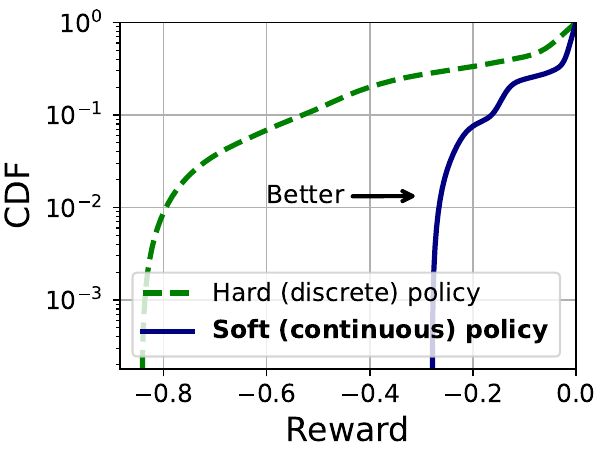}
    \caption{Comparison of the proposed hard and soft policies.}
    % \hz{change legend to be consistent with the text. Discrete .. and continuous ...}
    % \hz{change legend to "Hard (discrete) policy" and "Soft (continuous) policy".    If space is an issue, use "Hard policy" and "Soft policy"}
    \label{fig:policy_rep}
    \vspace{-0.15in}
\end{figure}

\textbf{Experimental Comparison.}
We conduct experiments on a 5G O-RAN testbed that comprises three cells and ten smartphones (see \S\ref{sec:implementation} for details). 
% \hz{please describe your experimental settings}. 
Each UE maintains a consistent position throughout the experiments while generating continuous data traffic demands. This controlled scenario allows us to isolate the impact of policy variations without mobility-induced effects. 
\mycut{The network operates in a synchronized manner, ensuring that all cells share the same frequency band, with DL and UL transmissions fully aligned.}
The hard and soft policy agents share the same model structure as described in \S\ref{subsec:diffusion}.
Fig.~\ref{fig:policy_rep} presents our measured reward.
It is evident that the soft policy representation significantly outperforms its hard counterpart, echoing the above analysis.

\subsection{Diffusion-Based Policy Generation}
\label{subsec:diffusion}

Based on the above policy representation, we propose a diffusion-based RL framework for online policy generation.

\textbf{Why Diffusion Model for RL?} 
Recently, diffusion-based RL has emerged as a strong candidate for the online decision-making process \cite{yang2023policy, chen2023boosting}.
It is well-suited for ICIM in O-RAN due to its ability to efficiently explore high-dimensional policy spaces while maintaining smooth and adaptive policy evolution. Unlike conventional RL approaches that may struggle with non-stationary interference patterns and dynamic resource constraints, diffusion-based RL leverages generative modeling techniques to learn a diverse distribution of optimal policies. 
This allows the policy agent to generalize across different network conditions and adapt to time-varying interference scenarios.

% Additionally, the decentralized nature of diffusion-based learning aligns well with O-RAN’s open architecture, enabling distributed decision-making among small cells and DUs while ensuring coordination through the RAN Intelligent Controller (RIC). By capturing the underlying structure of optimal interference mitigation strategies, diffusion-based RL can generate robust policies that balance spectrum efficiency, fairness, and network performance across multiple adjacent cells.

% Applying a diffusion model in double Q-learning has several appealing properties.
% \textbf{For our task, although the goal is to find the optimal reward value, the reality is that there are multiple feasible or optimal solutions to this optimization problem.} For instance,

\begin{figure}[t]
    \centering
        \centering
        \includegraphics[trim=80 50 180 50, clip, width=0.7\linewidth]{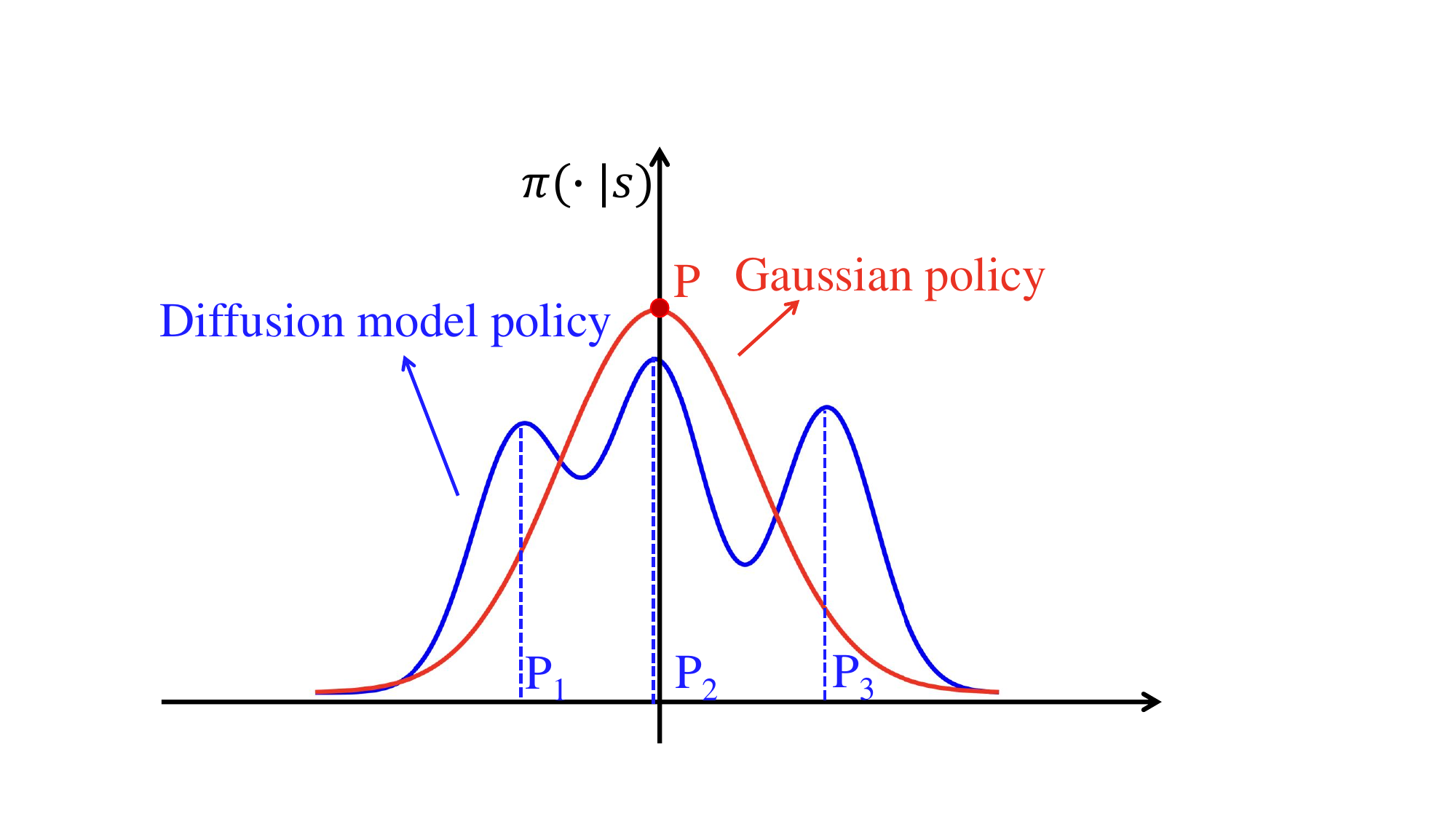}
        \caption{Gaussian modeling vs. diffusion modeling.}
        \label{fig:diffusionpolicy}
        \vspace{-0.15in}
\end{figure}

Consider two UEs, \( UE_1 \) and \( UE_2 \), for example. 
Each UE belongs to a different small cell. To mitigate inter-cell interference, multiple optimal resource allocation solutions may exist. For example, \( (UE_1 \!\!\to\!\! RB_1, UE_2 \!\!\to\!\! RB_2) \) is an optimal allocation; \( (UE_1 \!\!\to\!\! RB_2, UE_2 \!\!\to\!\! RB_1) \) is another optimal solution. As illustrated in Fig.~\ref{fig:diffusionpolicy}, the diffusion model is particularly well-suited for capturing the complex distribution of optimal actions.
It outperforms conventional Gaussian policies by enabling more accurate policy generation and offering faster convergence.

% where they transmit data in different frequency bands. However, an alternative solution could involve $UE_B$ using $BWG_1$ and $UE_A$ using $BWG_2$. We cannot directly specify the priority of spectrum usage by the base station because, in our experiment, we assume the base station is a central one rather than an edge base station. 
% Diffusion models are highly flexible and capable of capturing complex, multimodal distributions. This allows them to model situations where there are multiple potential outcomes or possibilities (e.g., different states), making them well-suited for handling problems (e.g. ICI), which may have multiple solutions.

\textbf{Diffusion-Based RL Framework.}
This framework takes the KPM and MAC data in Table~\ref{tab:kpmmacdata} as its input to generate $p(DU_k, UE_i, RGB_j)$ for all $k \in \mathcal{K}$, $i \in \mathcal{U}_k$, and $j \in \mathcal{J}$.
As shown in Fig.~\ref{fig:framework}, our RL framework integrates a diffusion model with double Q-learning networks to generate resource allocation policies (i.e., \textit{preference values}). The diffusion model acts as the policy generator, while the Q-learning networks serve as the evaluator and critic to ensure policy quality and stability. 
We train the diffusion model using the Q-values to progressively denoise the \textit{preference values}. At each denoising step, the model refines its output. We apply the principle of ``Guidance for Maximizing the Q-Function \cite{clifton2020q, watkins1992q},'' where the diffusion model prioritizes decisions that maximize the expected future reward (as indicated by the Q-value) during the optimization iterations. By adding this guidance at each iteration, the diffusion model is trained to maximize the reward.

\textbf{Double Q-learning.}
In the diffusion-based RL framework, we use a value-based model as the evaluator (critic), which generates values to guide the training of the diffusion model.  
Traditional Q-learning algorithms select the maximum target Q-value to update the current Q-value.
This may lead to overestimation of the Q-value due to noise and bias in data samples. 
To address this issue, we employ double-deep Q-learning networks to predict the Q-value of the policy generated by the diffusion model.
Of these two Q-networks, one predicts the current Q-values while the other predicts the target Q-value.

% It serves as the critic in our system, evaluating the value of the actions generated by the diffusion model.
% The Q-value function itself is learned in a conventional way. 
To enhance the learning stability and efficiency, following the design in \cite{fujimoto2018addressing}, we adopt two current Q-networks \( (Q_{\phi_{1}}, Q_{\phi_{2}}) \) for Q-value prediction and two target networks \( (Q_{\phi_{1}^{\prime}}, Q_{\phi_{2}^{\prime}}) \) to provide stable reference for current Q-value prediction. 
The smaller value from the two target networks is used to compute the target Q value.
This redundancy helps reduce Q-value overestimation. 
Denote $Q_{\phi_{i}^{\prime}}(\mathbf{s}_{t+1}, \mathbf{a}_{t+1}^{0})$  as the predicted value from the target Q network, where \(\mathbf{a}_{t+1}^{0} \sim \pi_{\theta^{\prime}} \).
Then, we compute the target Q-value as follows: 
\(y = r(\mathbf{s}_{t}, \mathbf{a}_{t}) + \gamma \min_{i=1,2} Q_{\phi_{i}^{\prime}}(\mathbf{s}_{t+1}, \mathbf{a}_{t+1}^{0})\), where \(r(\mathbf{s}_{t}, \mathbf{a}_{t})\) is the reward of taking action $\mathbf{a}_t$ at state $\mathbf{s}_t$ and $\gamma$ is the discount rate. 
Based on the target Q value, we optimize \(\phi_{i}\)---the weights in \( Q_{i} \)---by minimizing the below loss function:
\begin{equation}
   \mathcal{L}_Q(\theta) = \mathbb{E}_{(\mathbf{s}_{t}, \mathbf{a}_{t}, \mathbf{s}_{t+1}) \sim \mathcal{D}} \left[ \left( y - Q_{\phi_{i}}(\mathbf{s}_{t}, \mathbf{a}_{t}) \right)^{2} \right],
   \hfill 
   i \in \{1, 2\},
    \label{eq:q_loss}
\end{equation}
where $\mathcal{D}$ is the online dataset collected under policy $\pi_{\theta}$.

\textbf{Diffusion and Denoising Process.}
The training process of a diffusion model involves corrupting data with Gaussian noise at increasing levels and training a DNN to predict the added noise at each step. By minimizing the difference between the predicted and actual noise, the model implicitly learns the score function, enabling it to generate new samples by reversing the diffusion process.
The inference process of a diffusion model starts with pure Gaussian noise and iteratively denoises it using the trained model’s noise predictions. By reversing the diffusion process step by step, it gradually reconstructs a high-quality sample from the learned data distribution.
When using diffusion model within an RL framework, it should be stressed that there are two different timesteps:
one for diffusion noise-adding/denoising steps, and the other for RL iterations. 
We use superscript $ k \in  \{1, \dots, K\}$ to denote diffusion step index and subscript $t \in \{1, \dots, T\}$ to denote RL iteration index.
% \hz{Does one RF iteration have $K$ denoising steps in diffusion model?   Yes}
% with the covariance matrix fixed as \( \Sigma_{\theta}(\mathbf{a}^{i}, \mathbf{s}, i) = \beta_{i} \mathbf{I} \) and mean constructed as
% \[
% \mu_{\theta}(\mathbf{a}^{i}, \mathbf{s}, i) = \frac{1}{\sqrt{\alpha_{i}}} \left( \mathbf{a}^{i} - \frac{\beta_{i}}{\sqrt{1-\bar{\alpha}_{i}}} \epsilon_{\theta}(\mathbf{a}^{i}, \mathbf{s}, i) \right).
% \]

% The training process of diffusion model begins with an initial action \( \mathbf{a}_t^0 \), which undergoes a diffusion process. This diffusion process we add Gaussian noise to a at each step. The denoising process is implemented using a neural network, which refines the action estimation by incorporating the state \( s_t \) and progressing through multiple denoising steps over \( k \) iterations.

In RL, the policy/action must be generated based on the current system state.
This conditioning process is critical for policy generation.
Therefore, we define the conditional diffusion policy as follows:
\begin{equation}
    \pi_{\theta}(\mathbf{a} \mid \mathbf{s}) = p_{\theta}(\mathbf{a}^{0:K} \mid \mathbf{s}) = \mathcal{N}(\mathbf{a}^{K}; 0, \mathbf{I}) \prod_{k=1}^{K} p_{\theta}(\mathbf{a}^{k-1} \mid \mathbf{a}^{k}, \mathbf{s}),
    \label{eq: continu}
\end{equation}
where \(\mathbf{a}\) and \(\mathbf{s}\) are the action and state of an RL process, with $k$ denoting its step index. 
$\mathcal{N}(\cdot; \cdot, \cdot)$ denotes Gaussian noise following the given parameters.
Generally speaking, \( p_{\theta}(\mathbf{a}^{k-1} \mid \mathbf{a}^{k}, \mathbf{s}) \) could be modeled as a Gaussian distribution.

\begin{figure}[t]
        \centering
        \includegraphics[trim=20 20 20 40, clip, width=0.8\linewidth]{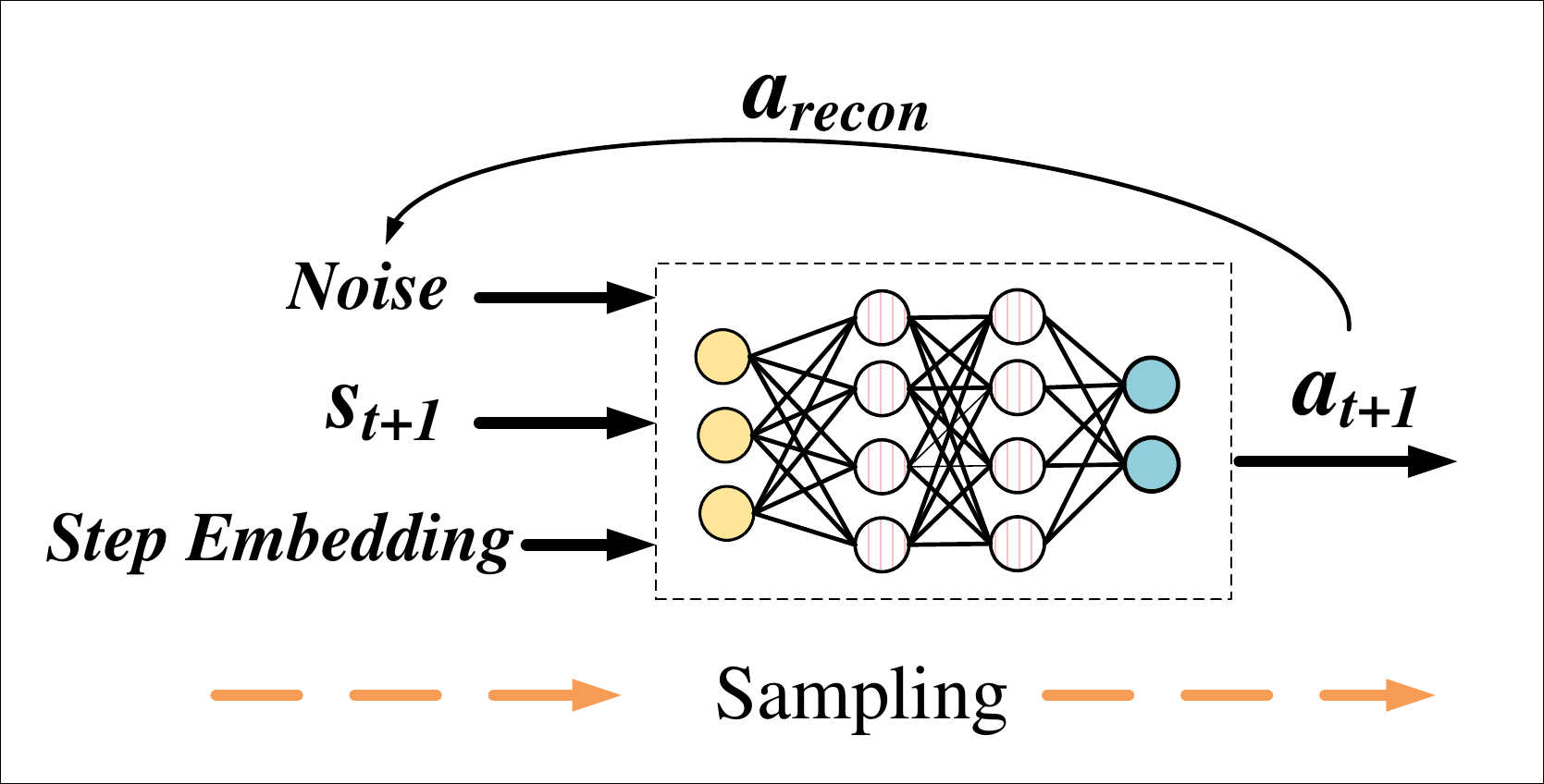}
        \caption{Iterative sampling process of diffusion model.}
        % \hz{Step index embeddings??? This input is a 16-dimensional step embedding, but I think directly writing the index in the figure is simpler and more intuitive.}
        \label{fig:Sampling}
        % \vspace{-0.15in}
\end{figure} 

To solve the multi-optimization problem, we follow \cite{wang2022diffusion} by parameterizing \( p_{\theta}(\mathbf{a}^{k-1} \mid \mathbf{a}^{k}, \mathbf{s}) \) as a conditional noise prediction model (see Fig.~\ref{fig:Sampling}).
Once the final action \( \mathbf{a}_t^K \) is obtained, it is used to determine the next action \( \mathbf{a}_{t+1} \). The \textit{preference values} update the objective function \( O(\mathbf{s}, \mathbf{a}) \), ensuring that the action selection process accounts for real-time network conditions. The optimization of the diffusion model is driven by a loss function, which balances denoising reconstruction and policy learning.

% This framework integrates diffusion-based policy generation with network-aware decision-making, enabling a more adaptive and robust action selection mechanism within O-RAN systems.

% We first sample \( a^N \sim \mathcal{N}(0, I) \) and then from the reverse diffusion chain parameterized by \( \theta \) as
% \[
% a^{i-1} \mid a^i = \frac{a^i}{\sqrt{\alpha_i}} - \frac{\beta_i}{\sqrt{\alpha_i (1 - \alpha_i)}} \epsilon_{\theta}(a^i, s, i) + \sqrt{\beta_i} \epsilon, \quad 
% \]
% , where \epsilon \sim \mathcal{N}(0, I), \quad \text{for } i = N, \dots, 1.
% Following DDPM \cite{ho2020denoising}, when \( i = 1 \), \( \epsilon \) is set as 0 to improve the sampling quality.

Following the Denoising Diffusion Probabilistic Model (DDPM) in \cite{ho2020denoising}, we train our conditional \(\epsilon\)-model---a Multi-Layer Perceptron (MLP) parameterized by $\theta$---based on the below loss function:
\[
\mathcal{L}_d(\theta) = \mathbb{E}_{k \sim U(\mathcal{K}), \epsilon \sim \mathcal{N}(0, I)} \left[ |\!|\epsilon - \epsilon_{\theta}(\sqrt{\bar{\alpha}_k} \mathbf{a} + \sqrt{1 - \bar{\alpha}_i} \mathbf{\epsilon}, \mathbf{s}, k)|\!|^2 \right],
\]
% , (\mathbf{s},\mathbf{a}) \sim \mathcal{D}
where \(\epsilon\) is the noise, \(U(\mathcal{K})\) is a uniform distribution over the discrete set as \(\{1, \dots, K\}\) and \(\mathcal{D}\) denotes the data samples in the database.

To improve learning efficiency, we inject the Q-value function from the Q-learning networks into the denoising process. 
Specifically, following the approach in \cite{wang2022diffusion}, we define 
$\mathcal{Q}(\theta) =   \frac{\mathbb{E}_{\mathbf{s} \sim \mathcal{D}}, a^0 \sim \pi_\theta \left[ Q(\mathbf{s}, \mathbf{a}^{0}) \right]}{\mathbb{E}_{(\mathbf{s}, \mathbf{a}) \sim \mathcal{D}} [|Q(\mathbf{s}, \mathbf{a})|]}$. 
Then, the loss function that we use to train the diffusion network is as follows:
\begin{equation}
    \mathcal{L}(\theta) = \mathcal{L}_{d}(\theta) - \eta \mathcal{Q}(\theta),
    %= \mathcal{L}_{d}(\theta) - \alpha \cdot \mathbb{E}_{\mathbf{s} \sim \mathcal{D}} \left[ Q(\mathbf{s}, \mathbf{a}^{0}) \right].
    \label{eq:diff_loss_combin}
\end{equation}
% \hz{please make sure it is "-" not "+". its -, we wanna minimize L, maximize Q-value}
% where $\beta$ is a coefficient to normalize the Q-value function, which appears to vary depending upon offline datasets.
% Following the approach in \cite{wang2022diffusion}, we set \( \beta \) to: 
% \begin{equation}
% \beta = \frac{\eta}{\mathbb{E}_{(\mathbf{s}, \mathbf{a}) \sim \mathcal{D}} [|Q(\mathbf{s}, \mathbf{a})|]}.
% \end{equation}
% \hz{can we combine (9) and (10)? So loss function is directly related to \(\eta\).}
where \( \eta \) is a hyperparameter that balances the two loss terms.
\(\eta\) plays a key role in model training. 
We will study it through experiments. 

\textbf{Sampling Process.}
Sampling refers to the process of generating new action (i.e., new policy for DUs' resource allocation) by reversing a noise-injection process, gradually refining random noise into a coherent output by following learned diffusion steps.
Fig.~\ref{fig:Sampling} illustrates our sampling process. 
Initially, the input includes randomly generated noise, state \( \mathbf{s}_{t+1} \), and the embedding of denoising step index \( k \).  The state information acts as a conditional guide, directing the network through iterative denoising processes to generate the action/policy $\mathbf{a}_{t+1}$ for state \( \mathbf{s}_{t+1} \).

% The state serves as a sampling condition, guiding the iterative denoising process, during which the network gradually refines the action representation and generates the final action \( \mathbf{a}_{t+1} \). 
% The sampling process requires the same number of steps as the denoising process in training. Additionally, the reconstructed action \( a_{\text{recon}} \) is used to facilitate stepwise sampling.

% As illustrated in Fig. \ref{fig:Sampling}, after multiple iterative denoising steps, the policy network progressively refines the action representation and ultimately generates a precise action output $\mathbf{a}_{t+1}$.
% Specifically, given a state condition, the policy network iteratively removes noise from the action representation to achieve accurate action reconstruction, utilizing the reconstructed action $\mathbf{a}_{recon}$ for iterative refinement. 

\textbf{Summary of Workflow.} The complete workflow of \pname is detailed in Alg.~\ref{alg:xdiff}. The parameter \(\rho\) controls the soft update of target networks using exponential moving average (EMA), ensuring a gradual adaptation to the latest policy while maintaining stability. A small \(\rho\) (e.g., 0.05) prevents abrupt changes. 
% reducing variance and improving convergence in reinforcement learning.

\begin{algorithm}[!t]
% \hz{this algorithm is not clear. It needs to be reorganized. Notation is messy. Done}
\caption{\pname Online Learning}
\label{alg:xdiff}
 % for ICIM \textit{preference value} Generation
% \hspace*{-\leftskip} \textbf{Initialize:} Policy network $\pi_\theta$, critic networks $Q_{v_1}$ and $Q_{v_2}$, target networks $\pi_{\theta^-}$, $Q_{v_1}^-$ and $Q_{v_2}^-$; 
\hspace*{-\leftskip} \textbf{Initialize:} Policy network $\pi_\theta$, Critic networks $Q_{\phi_{1}}, Q_{\phi_{2}}$, Target networks $\pi_{\theta^\prime}$, $Q_{\phi_{1}^\prime}$, $Q_{\phi_{2}^\prime}$ ;
%\noindent Initialize the dataset $\mathcal{D}_r \leftarrow \emptyset$

\hspace*{-\leftskip} \textbf{Input:} Action $\mathbf{a}_t$, Current State $\mathbf{s}_t$ and Next State $\mathbf{s}_{t+1}$
\hspace*{-\leftskip} \textbf{Output:} \textit{preference value} $\mathbf{a}_{t+1}$
\begin{algorithmic}[1]
% \State \textbf{Warm up:} %\hfill \textcolor{blue}{\textit{Initialize dataset with exploration}}
\State \textcolor{blue}{// \textit{Database initialization}} 
\State $\mathcal{D} \leftarrow \emptyset$
% \State
\For{$b \in \{0, \ldots, batch size\}$}
    \State Generate $\mathbf{a}_t^0 \sim \pi_{\theta^\prime}(\mathbf{a}_{t}|\mathbf{s}_{t})$ by Eq. (\ref{eq: continu}). 
    \State Play $\mathbf{a}_t^0$ and get $\mathbf{s}_{t+1} \sim \text{O-RAN Environment}$
    \State $\mathcal{D} \leftarrow \mathcal{D} \cup \{\mathbf{s}_t, \mathbf{a}_t^0, r_t, \mathbf{s}_{t+1}\}$   \hfill   
    % with exploration
\EndFor
%\State \textbf{For each iteration:} \hfill \textcolor{blue}{\textit{Iterative training and updating}}
% \State
\State \textcolor{blue}{// \textit{Online policy learning and generation}} 
\For{each iteration}  
% \hfill \textcolor{blue}{\(\triangleright\)\textit{Iterative training and updating}}
    % \State \textbf{Update the dataset:} \hfill \textcolor{blue}{\textit{Expand replay buffer}} % with new samples
    % \State Sample $\hat{a}_t^0 \sim \pi_\theta(a_t|s_t)$ by Eq. (\ref{eq: continu})
    % \State Play $\hat{a}_t^0$ and get $s_{t+1} \sim O-RAN environment$
    % \State $\mathcal{D} \leftarrow \mathcal{D} \cup \{s_t, \hat{a}_t, r_t, s_{t+1}\}$ 
    \State Repeat steps 4--6
    %\hfill \textcolor{blue}{\(\triangleright\)\textit{Expand replay buffer}} % with new samples
    % \State \textbf{Policy training:} \hfill \textcolor{blue}{\textit{Optimize policy using replay buffer}}
    \State Sample mini-batch $B = \{(\mathbf{s}_t, \mathbf{a}_t, r_t, \mathbf{s}_{t+1})\} \sim \mathcal{D}$  
    \State Sample $\mathbf{a}_{t+1}^0 \sim \pi_{\theta^\prime}(\mathbf{a}_{t+1}|\mathbf{s}_{t+1})$ by Eq. (\ref{eq: continu}).    
    % transition
    % \hfill \textcolor{blue}{\textit{Optimize policy using replay buffer}}
    % \State \textbf{Q-learning:} \hfill \textcolor{blue}{\textit{Update critic networks}}
    \State Update $Q_{\phi_{1}^\prime}$ and $Q_{\phi_{2}^\prime}$ based on Eq.~(\ref{eq:q_loss}) 
    % \hfill \textcolor{blue}{\(\triangleright\)\textit{Update Q networks}}
    % \State \textbf{Diffusion policy learning:} \hfill \textcolor{blue}{\textit{Minimize policy loss }}
    \State Update policy by minimizing Eq. (\ref{eq:diff_loss_combin})
    % \State \textbf{Update target networks:} \hfill \textcolor{blue}{\textit{Soft update target parameters}}
    \State $\theta^\prime \leftarrow \rho\theta + (1-\rho)\theta^\prime$ %\hfill \textcolor{blue}{\(\triangleright\)\textit{Soft update target weights}}
    \State $\phi_i^\prime \leftarrow \rho \phi_i + (1-\rho)\phi_i^\prime$ for $i \in \{1, 2\}$
\EndFor
\end{algorithmic}
\end{algorithm}

\subsection{Policy-Guided Resource Allocation at Individual DUs}

\label{subsec:pf}

\begin{figure} [!t]
    \centering
    \includegraphics[width=\linewidth, , trim= 15 3 30 3, clip]{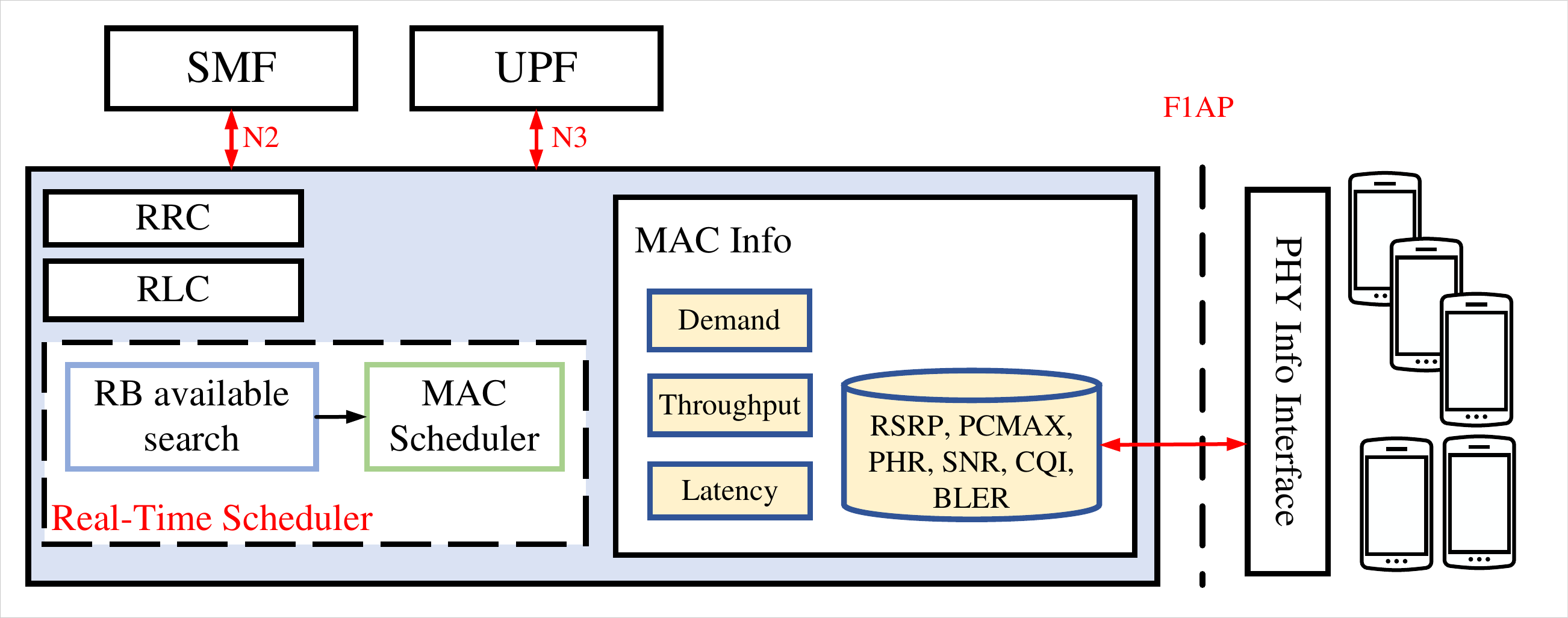}
    % \hz{reduce the height of this figure by moving the phones to the right side. Done}
    \caption{Architecture of 5G DU.}
    \label{fig:PFscheduler}
     \vspace{-0.15in}
\end{figure}

The policy generated by \pname must be taken by the DUs to guide their resource allocation.
Fig.~\ref{fig:PFscheduler} shows the architecture of an DU, where a real-time scheduler operates at the MAC layer for resource allocation and user scheduling.
In what follows, we first introduce the conventional UE scheduling algorithm in DU, and then explain how the policy generated by \pname is integrated into the existing scheduling algorithm.
\mycut{We note that we use ``resource allocation'' and ``user scheduling'' interchangeably.}

% MAC real-time scheduler for xDiff, highlighting its interaction with different network components. The scheduler operates within the MAC layer, which interacts with the Radio Resource Control (RRC) and Radio Link Control (RLC) layers. The real-time scheduler consists of an RBG available search module and a MAC scheduler, which dynamically allocates resources based on network conditions. The MAC information module collects key performance indicators, including demand, throughput, and latency, as well as critical radio metrics such as RSRP, PCMAX, PHR, SNR, CQI, and BLER. These metrics are exchanged via the F1AP interface, which connects the MAC layer to the PHY layer and the UE. Additionally, the MAC scheduler communicates with the Session Management Function (SMF) and User Plane Function (UPF) via the N2 and N3 interfaces, respectively, ensuring seamless coordination in resource allocation. This architecture enables efficient and adaptive scheduling to optimize network performance for \pname.

% \noindent\textbf{RBG Available Search.} 

\textbf{Proportional Fairness (PF) Scheduler.} 
PF is a popular scheduler that has been widely used in real-world cellular networks.
% It balances instantaneous channel conditions with the long-term throughput of each UE. 
% The scheduling process begins by traversing all UEs in the coverage area, where a PF coefficient is calculated for each UE. 
%Proportional Fairness (PF) scheduling algorithm assigns resources to users by balancing their instantaneous channel quality and long-term throughput.
It allocates resources to users based on a balance between maximizing throughput and ensuring fairness, by giving more resources to users with higher channel quality while preventing the system from favoring only the best users. It aims to optimize network performance while maintaining equitable access for all users.
Specifically, the PF scheduler calculates a metric for each user, which is a ratio of their instantaneous throughput (or CQI) to their historical throughput. This metric is used to decide which user should be allocated resources.
The metric for user $i$ at time slot $t$ is:
\begin{equation}
    \text{PF\_Metric}_i(t) = \frac{r_i(t)}{{R}_i(t-1)},
    \label{eq:pfmetric}
\end{equation}
where \( r_i(t) \) is the instantaneous achievable data rate of user \( i \) at time \( t \), which is predicted based on the CQI feedback. \( {R}_i(t) \) is the average historical throughput of user \( i \) in the previous time slot. 
%TBS????
The PF scheduler computes the PF metric for each user in each time slot, and selects the user with the highest PF metric for resource allocation.

% The PF scheduler selects the user \( i^* \) with the highest PF metric:

% \begin{equation}
% i^* = \arg\max_i PF_i(t).
% \label{eq:max_pf}
% \end{equation}

% This approach ensures that users with good channel conditions get scheduled more frequently while allowing users with poor channel conditions to still receive service over time.
% This factor is derived by considering the current MCS selection (reflecting instantaneous channel quality) and the historical throughput or realization rate of each UE, which helps maintain fairness over time. 
% Therefore, we have modified the scheduling strategy by introducing a transformation function, adjusting weight factors and incorporating additional constraints to better align with our system requirements.
% To improve the effectiveness of PF scheduling algorithm in real-time resource block allocation, we introduce a modification that incorporates a mapping coefficient, $coeff_{map}$, to better reflect the distribution of available RBs. 
% UEs are prioritized based on the ratio of their instantaneous data rate to their historical throughput, ensuring a balance between fairness and spectral efficiency. 

\textbf{Integrate Diffusion Policy into PF Scheduler.}
% However, this approach does not fully account for the variations in resource block grouping, which may impact the overall scheduling performance. Our preference value cannot be directly utilized in this framework. 
The MAC scheduler receives the policy (i.e., \textit{preference values}) generated by the diffusion model via the E2 interface and allocates RBs to UEs according to the policy. Specifically, for each UE, the scheduler aims to utilize the RBs with a high \textit{Preference Value} while avoiding those with a low \textit{Preference Value}. The scheduler achieves this objective through the below two steps.

\begin{itemize}
    
\item 
\textit{Step I: Adjust PF Metric.} 
Recall that $p(DU_k, UE_i, RB_j)$ indicates the $UE_k$'s preference for $RB_j$. 
A negative value means that this RB is not favorable due to the inter-cell interference.
We wish to exclude those RBs when calculating a UE's PF Metric.
To do so, we define a weight for UE $i$ by letting 
$w_{i} = \frac{\lvert \{ j \in \mathcal{J}: p(DU_k, UE_i, RB_j) < 0 \} \rvert}{N_{RB}}$.
It represents the percentage of favorable RBs for $UE_i$. 
Then, we incorporate the UE weight into the PF metric as follows.
\begin{equation}
    %PF_i= \frac{tbs_i(t)}{\sum_{r=1}^{t-1} th_i(r)} \times coff_{map}
    % PF_i(t)= \frac{tbs_i(t)}{th_i(t-1)} \times coeff_{map}
    \text{PF\_Metric}_i(t)= \frac{r_i(t)}{{R}_i(t-1)} \times w_i.
    \label{eq:tbs_pf}
\end{equation}
We note that, since DUs perform resource allocation independently, we omit DU index $k$ for simplicity.

\item 
\textit{Step II: RB Allocation.}  
The scheduler first prioritizes all UEs based on their PF\_Metric values and then allocates RB for each UE based on their priority. 
For each selected UE, the scheduler assigns RBs with the highest \textit{preference values} for transmission.
% continuing until the UE's demand is met or all RBs have been allocated.  
\mycut{
If the RBs within an allocated RB are not fully utilized by the current UE, the remaining RBs are retained and made available for scheduling to the next UE in priority order.
}
This allocation process continues iteratively until either the transmission demand of the current UE is fulfilled or all available RBs have been allocated.

\end{itemize}

\section{Implementation}
\label{sec:implementation}
% In this section, we present our implementation of \pname for evaluation.

\begin{figure} [t]
    \centering
    \includegraphics[trim=20 40 20 70, clip, width=\linewidth]{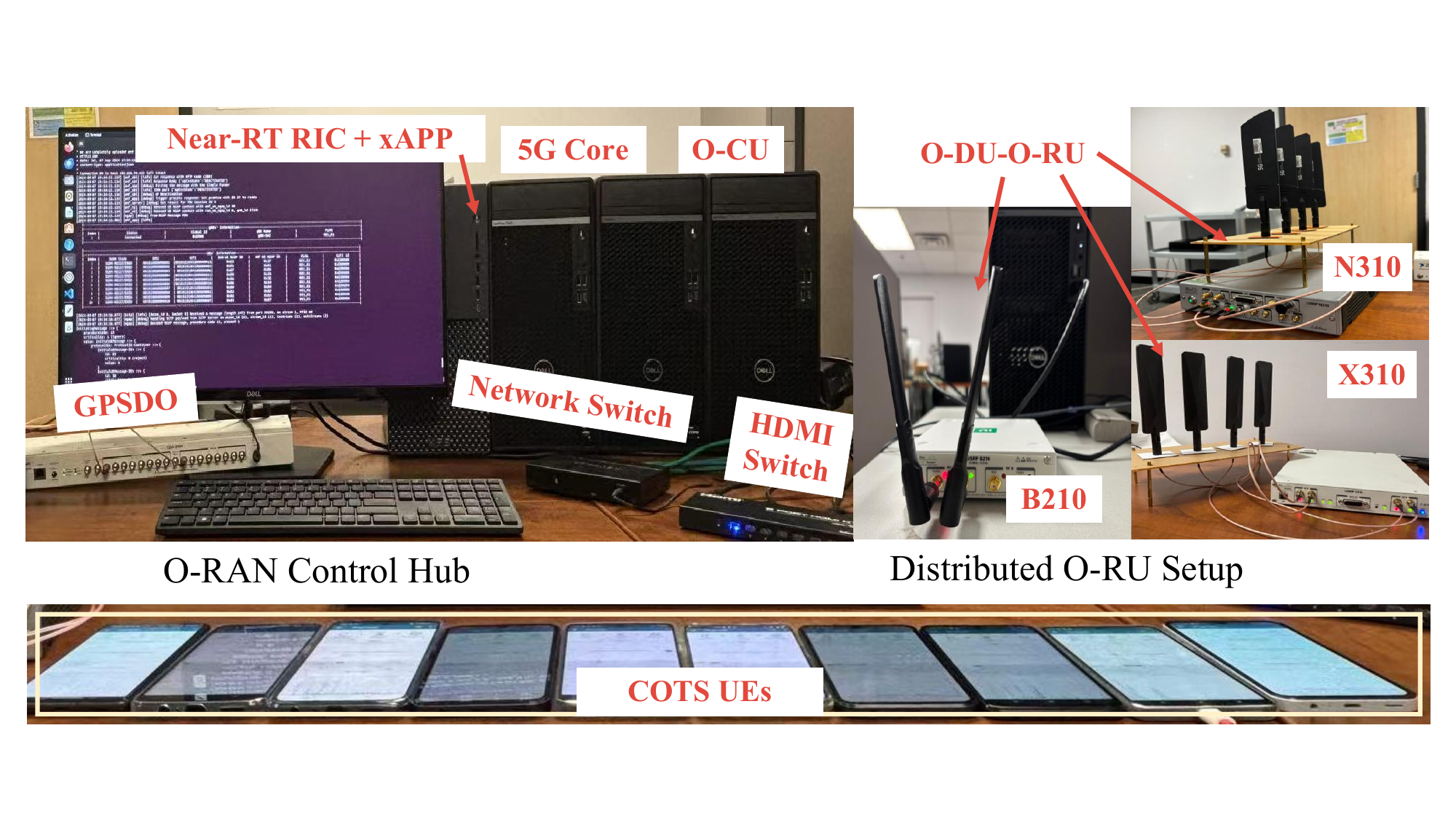}
    \caption{Testbed setup.}
    \label{fig:testbed}
    \vspace{-0.15in}
\end{figure}

% \begin{figure}[t]
%     \centering
%     \begin{minipage}{0.43\textwidth} 
%         \centering
%         \begin{subfigure}[b]{0.30\textwidth}
%             \centering
%             \includegraphics[trim=0 40 0 0, clip, width=\linewidth]{figures/B210.png}
%             \caption{USRP B210}
%             \label{fig:B210}
%         \end{subfigure}
%         \hspace{0\textwidth} 
%         \begin{subfigure}[b]{0.30\textwidth}
%             \centering
%             \includegraphics[trim=0 40 0 0, clip, width=\linewidth]{figures/X310.png}
%             \caption{USRP X310}
%             \label{fig:X310}
%         \end{subfigure}
%         \hspace{0\textwidth} 
%         \begin{subfigure}[b]{0.30\textwidth}
%             \centering
%             \includegraphics[trim=0 40 0 0, clip, width=\linewidth]{figures/N310.png}
%             \caption{USRP N310}
%             \label{fig:N310}
%         \end{subfigure}
%         \caption{DU and RU Setup}
%         \label{fig:testbedsetup}
%     \end{minipage}
%     \hz{combine it with Figure 10.}
% \end{figure}

\textbf{Testbed Setup.} 
Fig.~\ref{fig:testbed} illustrates our O-RAN experimental testbed, which consists of a 5G core network, one CU, three DUs, three RUs, Near-RT RIC, and ten commercial smartphones. 
The system operates on the n78 frequency band. 
The center frequency is 3319.68 MHz, and the subcarrier spacing is 30~kHz.
% We set up 3 O-DU-O-RUs on different host machines and different locations. 
% The figure also shows the deployment of three DU-RU units at different locations. We use five servers to implement the O-RAN system, hosting three O-DUs, one O-CU, a Near-RT RIC, and the 5G Core network. 
% The detailed hardware specifications are provided in Table~\ref{tab:testbed_hardware}. 
The testbed is configured so that all smartphones can access the Internet. 
% configuration of this testbed is designed to support internet access for COTS smartphones. 
% Local network connectivity is provided by a Netgear GS308v3 Ethernet switch, which interconnects the 5G core, O-CU, O-DU, and O-RU components. 
%
The three RUs were implemented using different USRP devices---N310, X310, and B210---to emulate the diversity of commercial RU equipment. 
Both N310 and X310 support 2x2 MIMO, while B210 supports 1x1 transmission only. 
The three USRP are synchronized in both frequency and time for TDD operations on the n78 band. 
% devices,  with the N310 supporting a 2x2 MIMO configuration and the X310 also supporting 2x2 MIMO. The USRP B210 device is used with a 1x1 antenna configuration. The USRP N310 is synchronized with an external GPSDO clock (CDA-2990) to improve clock precision. Additionally, the PPS signal is utilized for synchronization across the three DU units, ensuring simultaneous operation in either uplink or downlink mode. The radio units are tuned to the TDD n78 frequency band, centered at 3319.68 MHz, with a subcarrier spacing of 30 kHz. 
The ten smartphones are from various vendors, including Google Pixel, OnePlus, Motorola, Xiaomi, and Samsung.

\textbf{OpenAirInterface (OAI) Modifications.}
We use OAI \cite{openairinterface5g} 5G RAN for our experiments.
\mycut{Each DU and each CU have a unique ID.}
The DU then assigns a Radio Network Temporary Identifier (RNTI) to each of the newly admitted UEs as identification.
OAI only supports the PF algorithm, which cannot meet our requirements.
Therefore, we modified the downlink scheduler functions in 
\verb|oai/openair2/LAYER2/NR_MAC_gNB| and 
 the E2 interface in 
\verb|oai/openair2/E2AP/|\verb|RAN_FUNCTION|.
% In our setup, O-RU is only responsible for RF baseband processing because we use USRP as RU and USRP does not have the ability to be handled as a PHY layer.
% Additionally, in the Single Slice model, all sessions utilize a single BWP, where the BWP size corresponds to the total number of PRBs. 
% For \pname, we need to define a BWP for each slice, including its starting position and bandwidth size. The size can be obtained through \pname, while the starting position is determined by traversing through all of the PRBs.
% \noindent\textbf{5G Core, CU, DU, RU Split.}
\mycut{
We have extended 5G core OAICN \cite{oai_cn} for this project. The AMF manages user access, authentication, and mobility, while the UPF handles user data traffic routing and quality of service. 
}
%Our DU-RU uses the Split 8 scheme.

%\textcolor{red}{It includes O-CU and O-DU.  The fronthaul is cut on the right side (option 8), which means O-DU acts as a logical node hosting RLC, MAC, and High-PHY (including forward error correction, encoding/decoding, scrambling, modulation/demodulation) and Low-PHY (fast Fourier transforms, digital beamforming, and PRACH capture, filtering).  }

\textbf{Near-RT RIC.}
We use Mosaic5g Flexric \cite{flexric, flexran, foukas2016flexran} as our near-RT RIC. Flexric supports E2 Node agent, near-RT RIC, and xApp. 
It provides a flatbuffers encoding/decoding scheme as an alternative to ASN.1.
We use SWIG as an interface generator to enable C/C++ and Python development for xApps.
%Flexric supports E2AP versions v1.01/v2.03/v3.01 and KPM SM versions v2.01/v2.03/v3.00. 
We built our xApp with E2AP v2.03 and KPM v2.03.

% \begin{figure}
%     \centering
%     \begin{subfigure}{\linewidth}
%         \centering
%         \includegraphics[trim=30 150 20 80, clip, width=\linewidth, height=2cm]{experiments/motivation/spactr.pdf}
%         \caption{When BS1 is on the uplink, BS2 is on the downlink  without devices synchronization.}
%         \label{fig:freq-sinc}
%     \end{subfigure}
%     \begin{subfigure}{\linewidth}
%         \centering
%         \includegraphics[trim=30 150 20 80, clip, width=\linewidth, , height=2cm]{experiments/motivation/time_slot.pdf}
%         \caption{BS received IQ symbles after using PPS synchronization.}
%         \label{fig:iq-sinc}
%     \end{subfigure}
%     \caption{Comparison of with and without synchronization}
%     \Description{} 
%     \label{fig:sync_observation}
% \end{figure}

% 实验中有一个需要注意的问题，interference很大的时候，不只是ue会更换gnb，或者断开连接。受影响大的基gnb也会因为系统资源不够用而断开和xapp的通信，显示[E2 AGENT]: Condition not matched e.g., No UE matches condition. Emulator triggers this condition for testing, but not the RAN 

\textbf{\pname Implementation.}
We built an xAPP within the Near-RT RIC for our conditional diffusion model using a fully-connected DNN following the DDPM method in \cite{ho2020denoising}. 
% \textcolor{blue}{
To accelerate the online training process of the diffusion model and meet the timing requirements, we grouped the 106 RBs into 10 clusters for the diffusion model to generate the \textit{preference values}.
% }
We employed a DNN with a 4-layer MLP and mish activations for the diffusion model.
We use 256 hidden units for all layers. The input of $\epsilon(\theta)$ is the concatenation of the last step action vector, the current state vector, and the sinusoidal positional embedding of timestep $k$. The output of $\epsilon(\theta)$ is the predicted residual at diffusion timestep $K$.
For the double Q-learning networks, we adopt a similar MLP architecture as the diffusion policy; however, we use 4-layer MLPs with 256 hidden units.
%, as shown in  Fig.~\ref{fig:framework}.
% \hz{not sure if this should be placed here.  Here is OK}
For \pname, we need to define a BWP for each UE, including its starting position and bandwidth size. The starting position can be obtained through \pname, while the bandwidth size is determined by the \textit{preference values} and traversing through all the RBs.
% \subsection{Open-Source Code on GitHub}

\textbf{Open-Source Code:}
The source code of \pname is released on GitHub \cite{O-RANICIMPeihao}.

\section{Experimental Evaluation}
\label{sec:evaluation}

\begin{figure}[!t]
    \centering
    \begin{subfigure}{\linewidth}
        \centering
        \includegraphics[width=\linewidth, , trim= 10 3 10 15, clip]{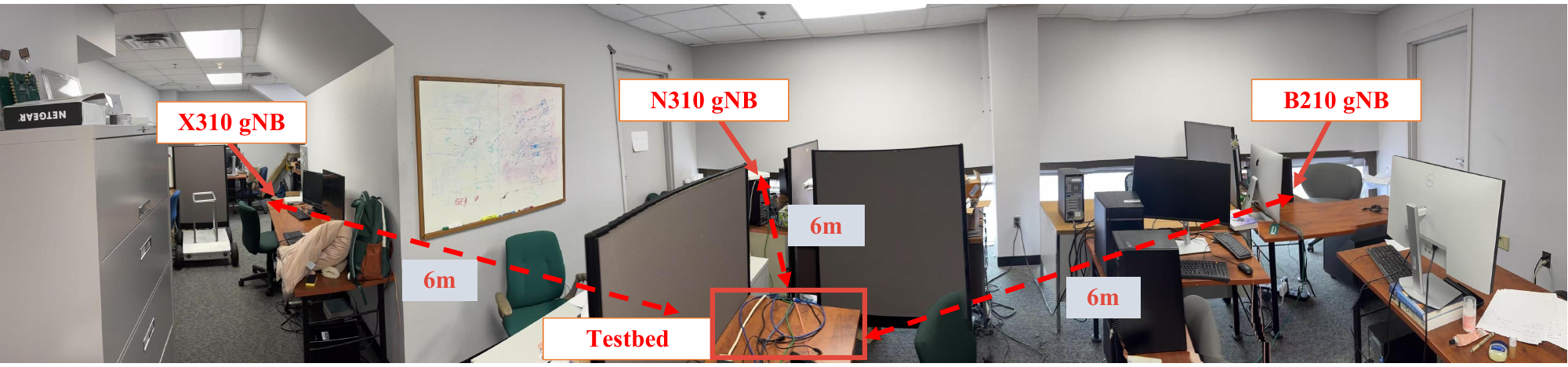}
        % \vspace{-0.1in}
        \caption{Lab scenario.}
        \label{fig:lab_senerio}
    \end{subfigure}
    \begin{subfigure}{\linewidth}
        \centering
        \includegraphics[trim= 80 3 100 5, clip, width=\linewidth]{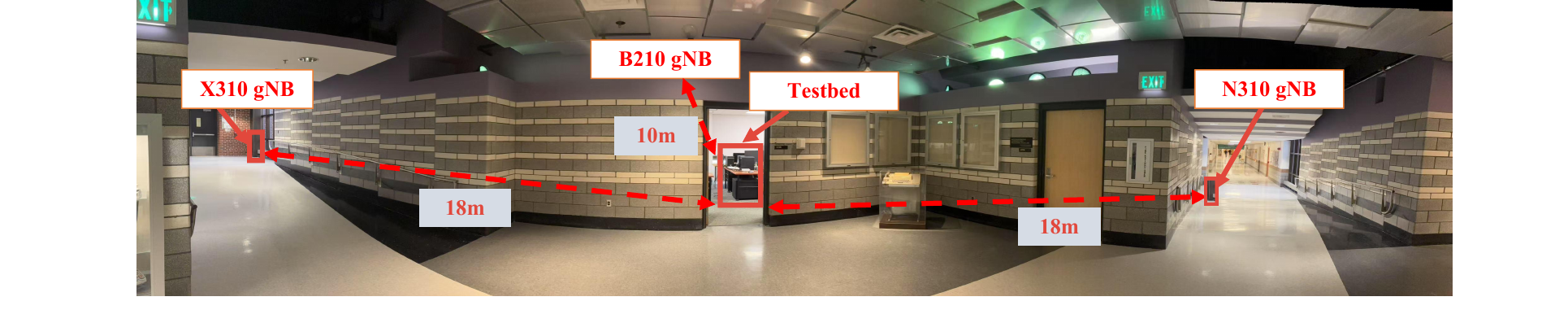}
        % \vspace{-0.15in}
        \caption{Building scenario.}
        \label{fig:building_senerio}
    \end{subfigure}
    \caption{Two scenarios for experimental evaluation.}
    % \Description{Experiment Scenarios.} 
    \label{fig:senerio}
    % \vspace{-0.15in}
\end{figure}

In this section, we evaluate the performance of \pname in lab and building scenarios as shown in Fig.~\ref{fig:senerio}. 
We use three metrics for the evaluation:  
throughput, delay, and the defined reward function. 
We note that the reward value is non-positive as it was defined to be the inverse of throughput/delay regret. 
We would like to answer the following questions through the evaluation.
\begin{itemize}
\item 
\textbf{Q1:}
How are KPM and MAC data affected by inter-cell interference?
This is important as \pname generates Near-RT ICIM policy based on KPM and MAC data. 

\item 
\textbf{Q2:}
Is diffusion model critical for \pname? 
What are the best values for its key parameters?

\item 
\textbf{Q3:}
How does \pname perform compared to the state-of-the-art ICIM solutions?
\end{itemize}

% \subsection{Evaluation Metrics and Baselines}

% To evaluate the performance of \pname, we will use metrics including per-UE throughput and latency distribution, as well as the defined reward value. 
% We note that the reward value is negative as it was defined to be the reverse of throughput/latency regret. 

\subsection{A Case Study}
%有一个实验需要解决的问题是：如果基站之间距离太大，干扰太小体现不出算法的有效性；如果基站之间的距离太小，干扰太大容易造成用户断联和xapp断开连接的问题，中间断开就没法完整展示

%没有办法采用CQI等作为INR的原因是 这些参数除了受interference影响还受到环境影响很大

% Consider a system consisting of three base stations and three UEs, where each base station is connected to one UE. We analyze the impact of inter-cell interference at three different levels: the physical layer, the MAC layer, and the application layer. At the physical layer, we focus on the transmission process and investigate how interference affects signal strength and overall system throughput. At the MAC layer, we explore how resource scheduling policies affect the distribution of interference and its subsequent impact on network performance. At the application layer, we evaluate the impact of interference on key performance metrics such as latency and throughput, as well as the case of different demands, to gain a comprehensive understanding of how interference affects multilayer systems.

% 这里我们可以注意到，并不是ph越大，rsrp越大

% 根据比较，n310的系统容量是最大的，吞吐量最高，延迟最小

% \textbf{PHY-Layer Analysis.}
% In-sync PH represents the power of the synchronization signal received by the UE from the base station. A higher value indicates better synchronization and a more stable connection. PCMAX which means maximum allowable transmission power for the UE in the uplink. It helps control power levels to avoid interference and optimize network performance. Average RSRP is the average power of the reference signal received from the base station. A higher RSRP indicates a stronger downlink signal, reflecting better connection quality.

We consider a simple network in the lab scenario as a case study to examine the operations of \pname. 

% While higher in-sync PH values generally indicate stronger transmission synchronization, they do not necessarily correlate with higher RSRP. This discrepancy is influenced by the inherent performance differences between USRP models and the varying channel conditions experienced by each UE, which affect signal propagation and reception quality.

\textbf{Two-Cell Case.}
We first consider a network with two cells, each serving a single smartphone. The core network generates persistent data traffic for the two smartphones at 50 Mbps and 60 Mbps, respectively. We aim to observe the MAC-layer data and evaluate its performance under inter-cell interference. To do this, we first activate smartphone 1 at time moment \(T_1\), followed by smartphone 2 at time moment \(T_2\). We collect smartphone 1's MAC-layer data at its serving DU, including its Power Headroom (PHR), Channel Quality Indicator (CQI), Modulation and Coding Scheme (MCS), BLER, throughput (TP), and queueing delay.

\begin{figure} [!t]
    \centering
    \includegraphics[width=\linewidth, , trim= 40 0 20 0, clip]{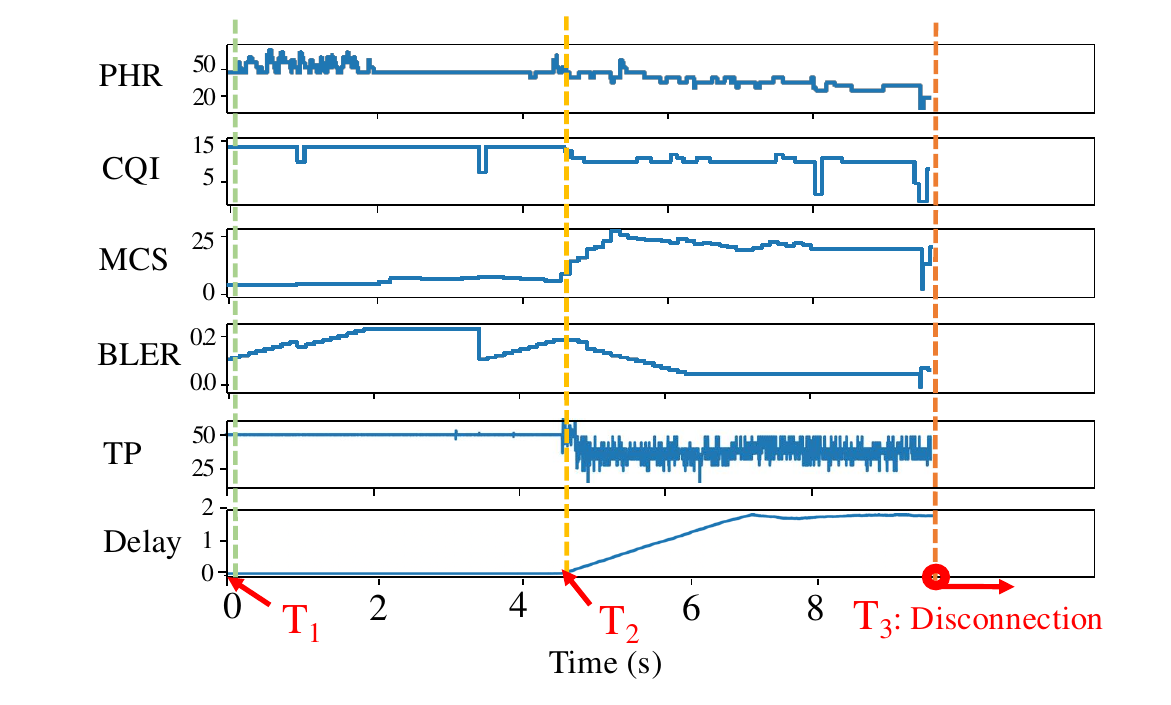}
    \caption{MAC-layer observations. \(UE_1\) and \(UE_2\) start their downloading requests at 50 Mbps at $T_1$ and $T_2$, respectively.
    Throughput (TP) was measured in Mbps, and delay was measured in seconds.}
    \label{fig:CQI}
\end{figure}

Fig.~\ref{fig:CQI} presents our measurement data, with the time moments \(T_1\) and \(T_2\) marked along the x-axis. We make the following observations. Between \(T_1\) and \(T_2\), since only one smartphone is active, there is no inter-cell interference. As a result, smartphone 1 achieves ideal connection performance. During this period, it maintains a stable link with high PHR values of around 50, high CQI values of around 15, and consistent MCS values. Its real throughput remains stable at 50 Mbps, consistently meeting its demand. The queueing delay stays low at around a few milliseconds.

From time moment \(T_2\), the network has two active smartphones with aggressive traffic demands, generating significant inter-cell interference. The impact of interference is reflected in smartphone 1's MAC data.
As shown in Fig.~\ref{fig:CQI}, since time moment \(T_2\), smartphone 1 experiences a considerable performance drop, including decreased PHR, declined CQI, reduced throughput, and increased delay. At time moment \(T_3\), smartphone 1 disconnects from the network due to the overwhelming interference.

These observations confirm the underlying relationship between inter-cell interference and MAC data, supporting our design of the diffusion model that generates ICIM policies based on MAC/KPM observations.

\begin{table}[!t]
    \centering
    \small
    \caption{RU power parameters.}
    \resizebox{0.38\textwidth}{!}{
        \begin{tabular}{|c|c|c|c|c|}
             \hline
            & USRP & \makecell{in-sync PH \\ (dB)} & \makecell{PCMAX \\ (dBm)} & \makecell{Average RSRP \\ (dBm)} \\ 
            \hline
            $UE_1 \in gNB_1$ & B210 & 55 & 17 & -88 \\ 
            \hline
            $UE_2 \in gNB_2$ & N310  & 42 & 21 & -73 \\ 
            \hline
            $UE_3 \in gNB_3$ & X310 & 32 & 17 & -89 \\ 
            \hline
        \end{tabular}
    }
    \label{tab:usrp_para}
    % \vspace{-0.1in}
\end{table}

\textbf{Three-Cell Case. }
We now examine the impact of interference in a three-cell network, where each cell serves a single UE (smartphone).  
Table~\ref{tab:usrp_para} presents the key measured parameters of the three UEs, including their in-sync PHR, maximum transmission power (PCMAX), and average Received Signal Reference Power (RSRP). These measurements reflect the link quality of each UE.  
For each UE in each cell, we gradually increased its throughput demand from 0 to 80 Mbps using the \verb|iperf| and measured its throughput and delay performance in the network with and without \pname.

\begin{figure}[!t]
    \centering
    \begin{subfigure}[b]{0.48\columnwidth}
        \centering
        \includegraphics[trim=0 0 10 0, clip, width=\linewidth]{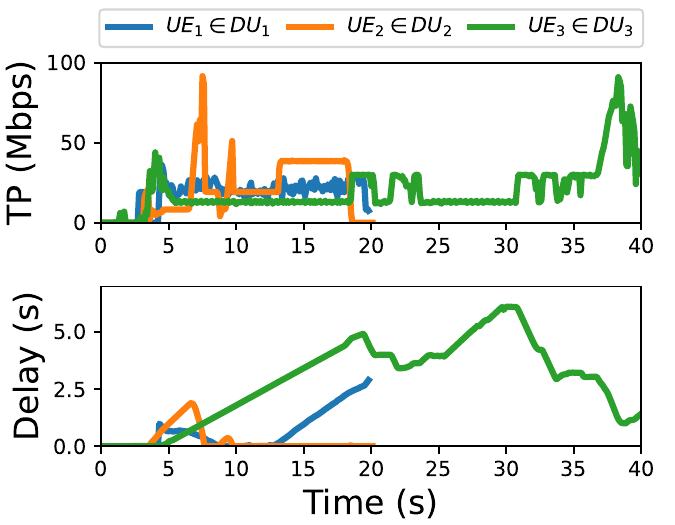}
        \caption{W/o \pname.}
        \label{fig:individual_method}
    \end{subfigure}
    \begin{subfigure}[b]{0.48\columnwidth}
        \centering
        \includegraphics[trim=0 0 10 0, clip, width=\linewidth]{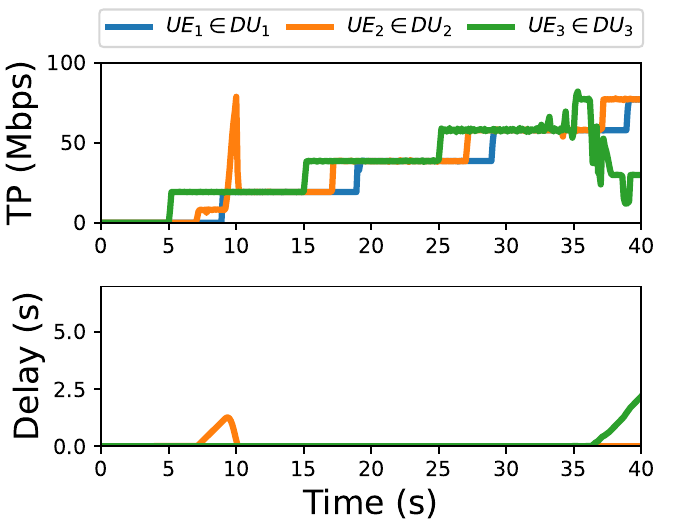}
        \caption{W/ \pname.}
        \label{fig:xdiff_method}
    \end{subfigure}
    % \hz{current legend blocks the curves. Place the legend on top of the figures in one row.}
    \caption{Performance comparison of the network w/ and w/o \pname.}
    \label{fig:testcases_withici}
    % \vspace{-0.15in}
\end{figure}

Fig.~\ref{fig:testcases_withici} presents an instance of our measurement results. We make the following observations.  
When the three network cells allocate resources independently (i.e., without \pname), the throughput of the three UEs fluctuates dramatically and frequently falls below their throughput demands. This is due to the independent resource allocation at each cell, which leads to dynamic inter-cell interference.  
Compared to throughput, the impact of interference on communication delay is even more significant. The delay for one UE spikes to as high as 5 seconds.  
More critically, inter-cell interference causes UE disconnections when each UE continues to increase its throughput demand. As shown in the figure, \(UE_1\) and \(UE_2\) lose connection at the 20-second mark due to severe interference.  
In contrast, the network with \pname exhibits much more stable throughput and delay performance. Moreover, the network meets the throughput and delay demands of the three UEs most of the time, and no UE disconnections are observed.  
This demonstrates the effectiveness of \pname in ICIM.

\begin{figure}
    \centering
    \begin{subfigure}[b]{0.32\columnwidth}
        \centering
        \includegraphics[trim=0 0 10 5, clip, width=\linewidth]{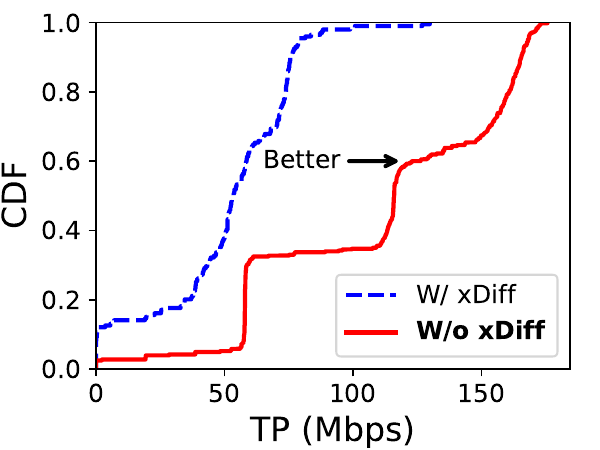}
        \caption{Throughput}
        \label{fig:tp_case}
    \end{subfigure}
    \begin{subfigure}[b]{0.32\columnwidth}
        \centering
        \includegraphics[trim=0 0 10 5, clip, width=\linewidth]{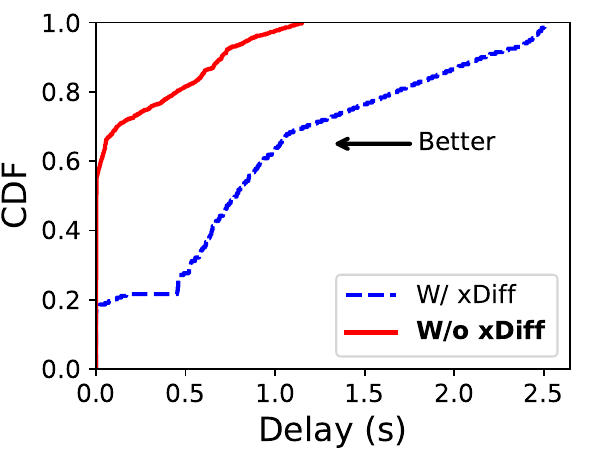}
        \caption{Delay}
        \label{fig:latency_case}
    \end{subfigure}
    \begin{subfigure}[b]{0.32\columnwidth}
        \centering
        \includegraphics[trim=0 0 10 5, clip, width=\linewidth]{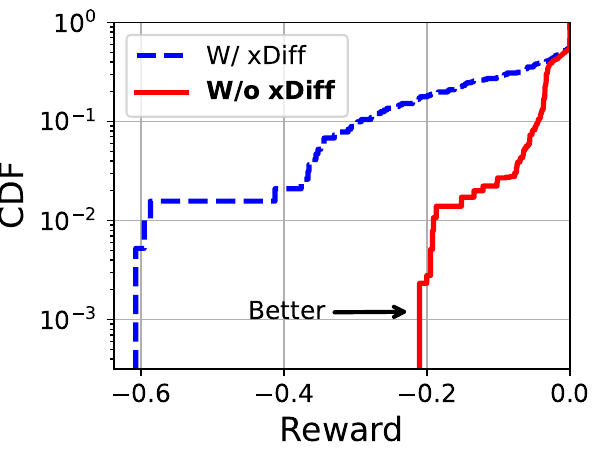}
        \caption{Reward}
        \label{fig:reward_case}
    \end{subfigure}    
    \caption{Measured performance in the case study.}
    % \hz{change the legend to: W/ xDiff and W/o xDiff.   Done}
    \label{fig:compare_xdiffandindividual}
    % \vspace{-0.15in}
\end{figure}

We repeated the above measurements multiple times, placing the three smartphones in different locations to collect their throughput and delay data.  
Fig.~\ref{fig:compare_xdiffandindividual} presents our measurement results.  
It is evident that using \pname significantly improves the network's throughput, delay, and reward performance.  
This further confirms the effectiveness of \pname in ICIM.

% the scenario without any ICIM demonstrates significant fluctuations in throughput among \(UE_1\), \(UE_2\), and \(UE_3\), accompanied by noticeable delays. 
% The delay for \(UE_3 \in DU_3\) increases sharply after \(t = 20 s\), and \(UE_1\), \(UE_2\) experience disconnections with the system. 
% In contrast, Fig.~\ref{fig:xdiff_method} illustrates \pname’s ability to maintain stable throughput and reduce delay. Throughput variations are minimized across all UEs, and latency is significantly suppressed. 
% Moreover, interference is mitigated, and the connections of the UEs remain stable throughout the testing period.
% These results confirm that \pname effectively manages interference and enhances resource allocation.

% We compared the performance of \pname against that scenario without any ICIM using cumulative distribution functions (CDFs) for reward, throughput, and latency, as shown in Fig.~\ref{fig:compare_xdiffandindividual}. In Fig.~\ref{fig:reward_case}, \pname significantly outperforms the baseline by achieving higher rewards. 
% For throughput (Fig.~\ref{fig:tp_case}), \pname consistently achieves higher values, demonstrating superior data transmission efficiency. Fig.~\ref{fig:latency_case} shows that \pname has notably lower latency than the baseline. Collectively, these results validate the effectiveness of \pname in enhancing network performance across multiple metrics.

\subsection{Ablation and Parameter Studies}

\begin{figure}
    \centering
        \begin{subfigure}[b]{0.32\columnwidth}
            \centering
            \includegraphics[trim=30 15 30 10, clip, width=\linewidth]{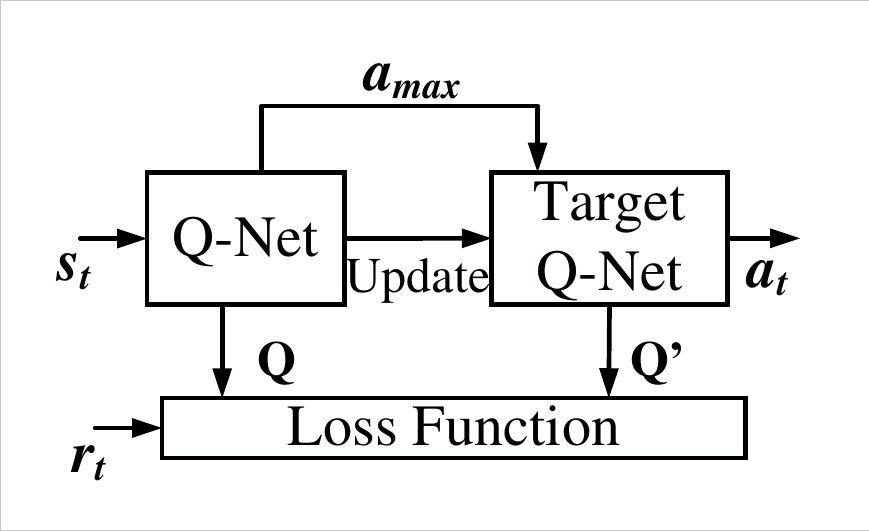}
            \caption{Double DQN.}
            \label{fig:doubleQN}
        \end{subfigure}
        \hspace{0\textwidth} 
        \begin{subfigure}[b]{0.32\columnwidth}
            \centering
            \includegraphics[trim=25 10 25 10, clip, width=\linewidth]{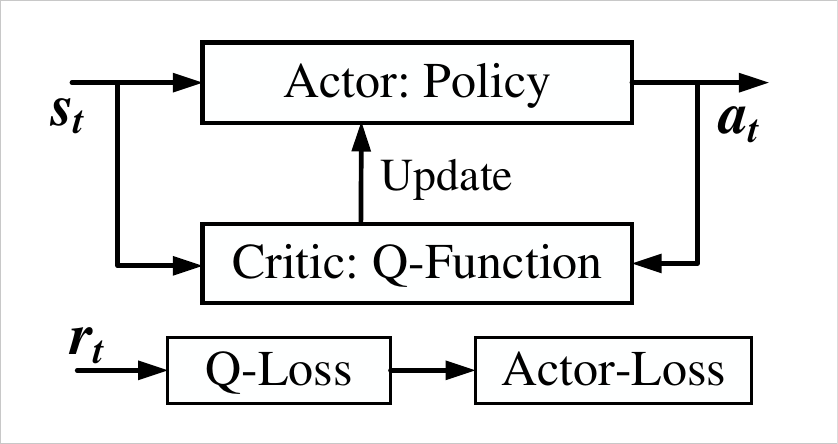}
            \caption{DDPG.}
            \label{fig:ddpg}
        \end{subfigure}
        \hspace{0\textwidth} 
        \begin{subfigure}[b]{0.31\columnwidth}
            \centering
            \includegraphics[trim=30 10 25 10, clip, width=\linewidth]{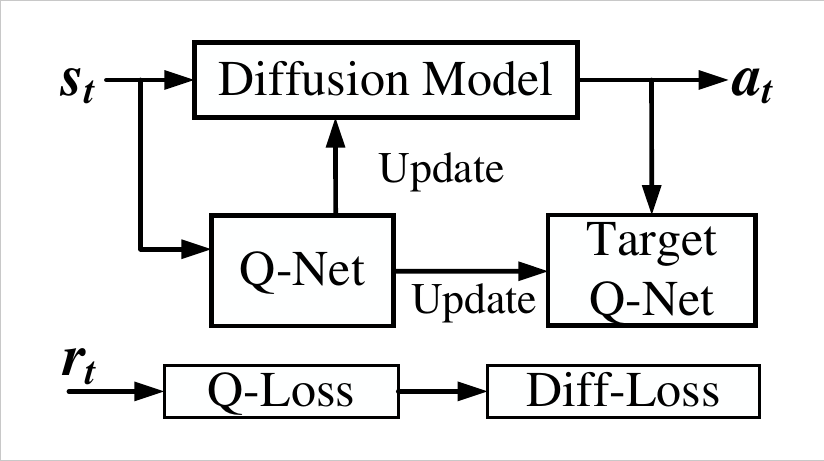}
            \caption{\pname.}
            \label{fig:xdiff}
        \end{subfigure}
        \caption{Comparison of Different RL Algorithms}
        % \hz{squeeze the text to enlarge the figures. Make sure the text is legible. Done}
        \label{fig:compare_rlmodel}
        % \vspace{-0.15in}
\end{figure}

\textbf{Ablation Study.}
\pname consists of multiple key components, including a diffusion model and Q-networks. To ensure that the performance gain in \pname is primarily attributed to the diffusion model rather than other components, we conduct an ablation study. Specifically, we compare \pname with two variant models obtained by removing its diffusion components. Fig.~\ref{fig:compare_rlmodel} highlights the network architecture comparison between \pname and its two counterpart models, which we describe as follows.

% In this section, we conduct an ablation study to evaluate the effectiveness of using a diffusion model as a policy. Specifically, we investigate why the diffusion model outperforms traditional methods in this context. 
% To validate our approach, we compare the performance of our method, which integrates the diffusion model as the policy, with two baseline algorithms: Double Q-learning and standard DDPG (Deep Deterministic Policy Gradient) as shown in Fig. \ref{fig:compare_rlmodel}, neither of which utilize diffusion models.

% Through systematic analysis of the results, we demonstrate that the diffusion model consistently outperforms traditional methods, offering superior performance in policy optimization.
% Fig. \ref{fig:compare_rlmodel} compares different online RL algorithms, including Double DQN (value based), DDPG (policy based), and our proposed \pname method. 

% 比较三个，1. 只用q-learning， 2. 采用actor- critic，其中critic用q- learning，actor用dnn+高斯采样?这个直接换成DOUBLE Q-LEARNING吧 3. xdiff

\begin{itemize}

    \item \textbf{Double Deep Q-learning Algorithm (DDQN):} 
    We remove the diffusion model in \pname to create this DDQN model. It uses a Q-network to generate the actions and another Q-network to predict the target function values. Similar to \pname, it employs double Q-networks to mitigate overestimation bias. The target Q-network is updated based on the next state and the loss function is computed using the reward signal.

    \item \textbf{Deep Deterministic Policy Gradient (DDPG):} 
    DDPG is a specialized Actor-Critic RL method designed for continuous action spaces.
    It is a model-free actor-critic framework, consisting of an actor network implemented as an MLP and a critic network that evaluates the Q-function.
    It does not have a diffusion model but also uses two Q-networks to improve stability. 
    % \item \textbf{\pname: } Our proposed \pname 
    % % integrates a diffusion model into the RL framework, enhancing action sampling by leveraging learned distributions. Unlike conventional RL methods, xDiff 
    % refines action generation through a diffusion-based model, which is updated using Q-network feedback and a dedicated diffusion loss. 
\end{itemize}

% This structure enables more robust exploration and improves decision-making by effectively learning complex action distributions.

\begin{figure}[!t]
    \centering
        \begin{subfigure}[b]{0.47\columnwidth}
            \centering
            \includegraphics[trim=0 0 0 0, clip, width=\linewidth]{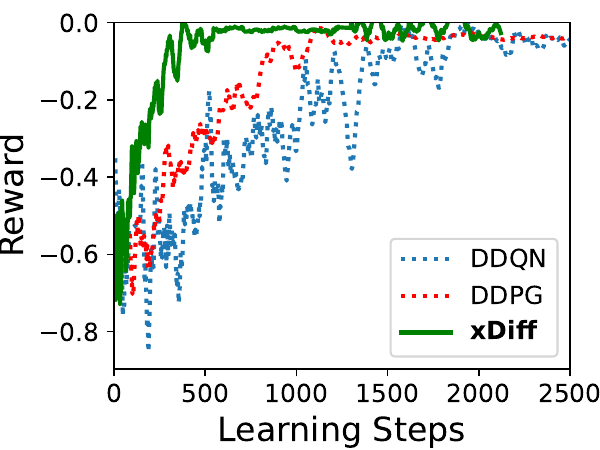}
            \caption{Convergence.}
            \label{fig:converge_compare}
        \end{subfigure}
        \hspace{0\textwidth} 
        \begin{subfigure}[b]{0.47\columnwidth}
            \centering
            \includegraphics[trim=0 0 0 0, clip, width=\linewidth]{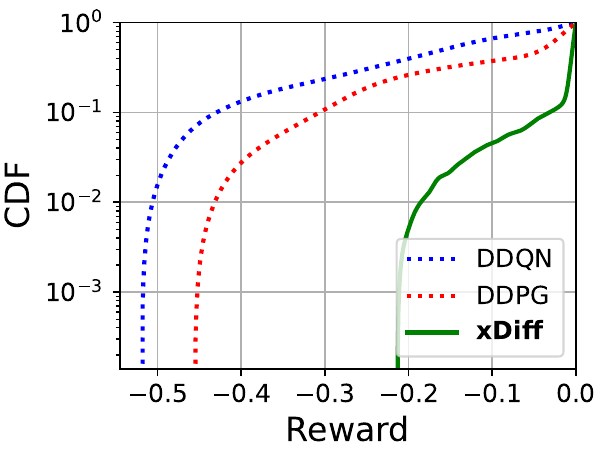}
            \caption{Reward.}
            \label{fig:aicompare}
        \end{subfigure}
        \caption{Experimental results of ablation study.}
        % \hz{left figure: y lim should be up to 0. Legend: change "Double Q" to "DDQN".}
        \label{fig:diffusion_compare}
        % \vspace{-0.15in}
\end{figure}

We implemented these three policy generation models on our testbed and evaluated their performance under identical scenarios.  
The experiments involved three cells and ten smartphones.  
Fig.~\ref{fig:diffusion_compare} presents our experimental results in terms of convergence speed and reward distribution.  
Fig.~\ref{fig:converge_compare} shows that \pname achieves the fastest convergence speed when there is a change in the network state (e.g., increased UE throughput demands).  
Fig.~\ref{fig:aicompare} demonstrates that \pname attains the highest reward compared to its two counterparts.  
These observations confirm that \pname outperforms its ablated counterparts and highlight the critical role of the diffusion model in generating efficient policies.

% compares the performance of the \pname, against two baseline algorithms, in terms of convergence behavior and reward distribution under ICI conditions. Fig.~\ref{fig:converge_compare} demonstrates that \pname achieves faster convergence and consistently attains higher reward values over learning steps. Fig.~\ref{fig:aicompare} further highlights the superiority of \pname, which indicates significantly improved reward outcomes. 

\begin{figure}[t]
    \centering
    \begin{subfigure}[b]{0.23\textwidth}
        \centering
        \includegraphics[trim=0 0 0 0, clip, width=\linewidth]{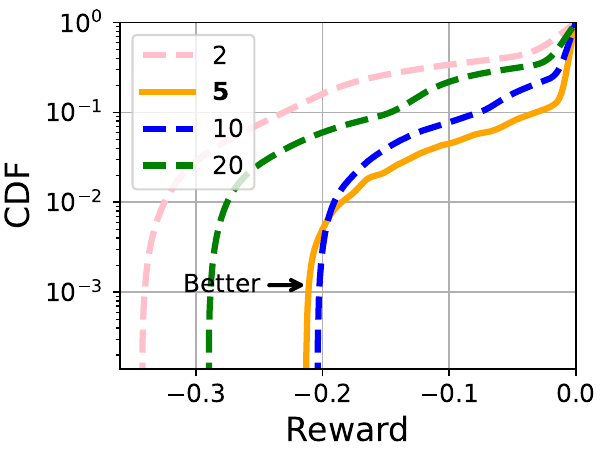}
        \caption{\# of denoising steps, i.e., $N$.}
        \label{fig:ncompare}
    \end{subfigure}
    \begin{subfigure}[b]{0.23\textwidth}
        \centering
        \includegraphics[trim=0 0 0 0, clip, width=\linewidth]{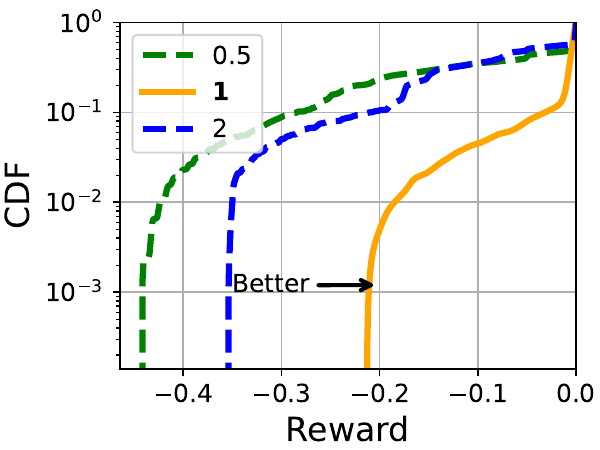}
        \caption{Loss-balance parameter $\eta$.}
        \label{fig:ucompare}
    \end{subfigure}
    \caption{Ablation study of hyperparameters}
    \label{fig:hyperparameters}
    % \vspace{-0.2in}
\end{figure}

\begin{figure*}
    \centering
    % \hspace{0\textwidth}  % 控制图片之间的间隔
    \begin{subfigure}[b]{0.24\textwidth}
        \centering
        \includegraphics[trim=0 0 0 0, clip, width=\linewidth]{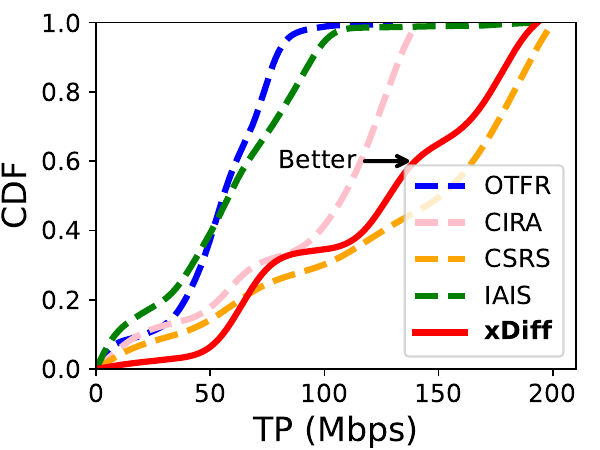}
        \caption{Throughput.}
        \label{fig:lab_thp}
    \end{subfigure}
    % \hspace{0\textwidth}
    \begin{subfigure}[b]{0.24\textwidth}
        \centering
        \includegraphics[trim=0 0 0 0, clip, width=\linewidth]{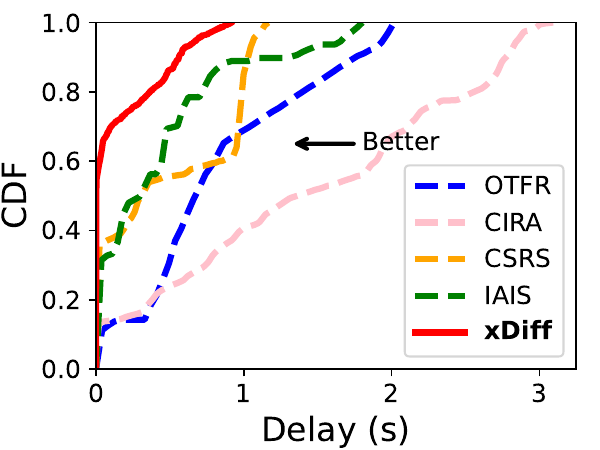}
        \caption{Delay.}
        \label{fig:lab_latency}
    \end{subfigure}
    % \hspace{0\textwidth}
    \begin{subfigure}[b]{0.24\textwidth}
        \centering
        \includegraphics[trim=0 0 0 0, clip, width=\linewidth]{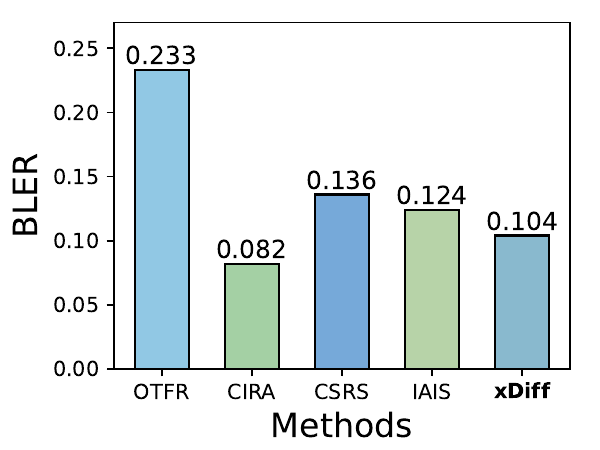}
        \caption{Avg BLER.}
        \label{fig:lab_bler}
    \end{subfigure}
    \begin{subfigure}[b]{0.24\textwidth}
        \centering
        \includegraphics[trim=0 0 0 0, clip, width=\linewidth]{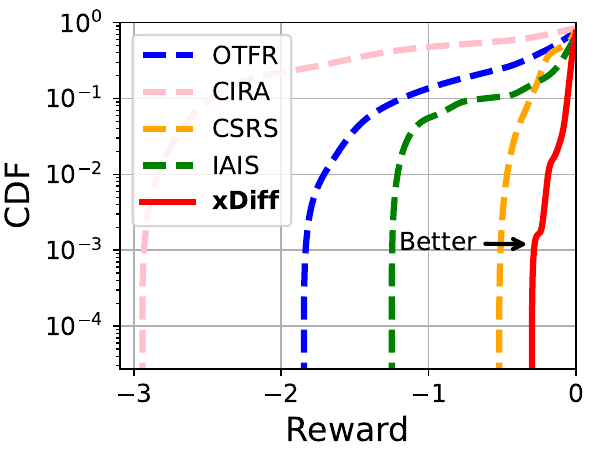}
        \caption{Reward.}
        \label{fig:lab_reward}
    \end{subfigure}    
    \caption{Performance comparison of \pname and existing approaches in lab scenario}
    \label{fig:compare_lab}
\end{figure*}

\begin{figure*}[!t]
    \centering
    % \hspace{0\textwidth}  % 控制图片之间的间隔
    \begin{subfigure}[b]{0.24\textwidth}
        \centering
        \includegraphics[trim=0 0 0 0, clip, width=\linewidth]{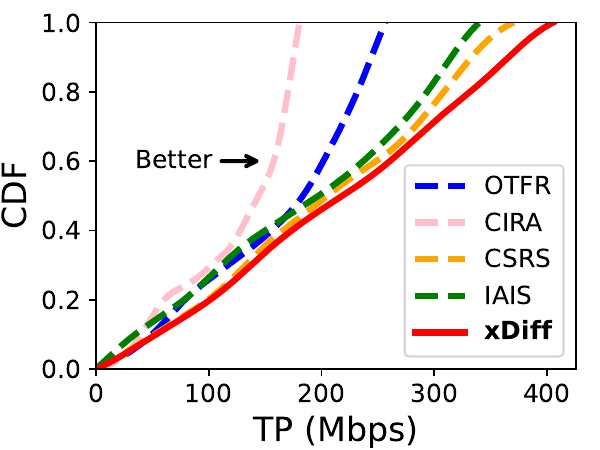}
        \caption{Throughput.}
        \label{fig:building_thp}
    \end{subfigure}
    % \hspace{0\textwidth}
    \begin{subfigure}[b]{0.24\textwidth}
        \centering
        \includegraphics[trim=0 0 0 0, clip, width=\linewidth]{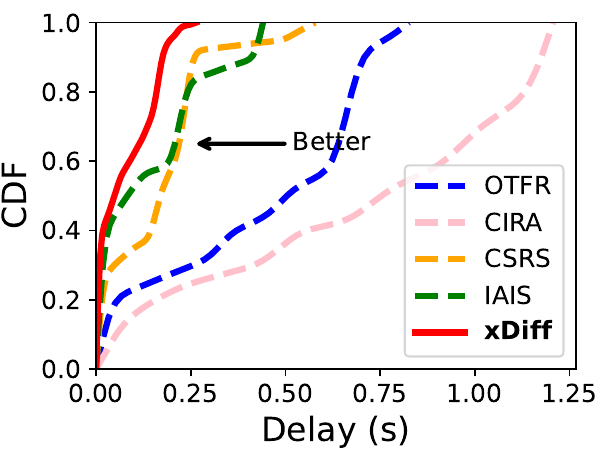}
        \caption{Delay.}
        \label{fig:building_latency}
    \end{subfigure}
    % \hspace{0\textwidth}
    \begin{subfigure}[b]{0.24\textwidth}
        \centering
        \includegraphics[trim=0 0 0 0, clip, width=\linewidth]{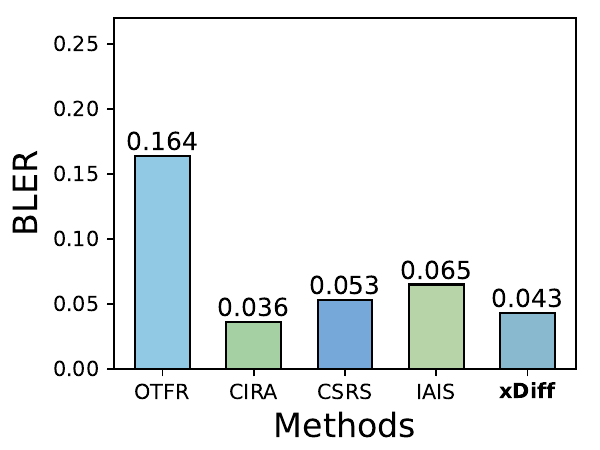}
        \caption{Avg BLER.}
        \label{fig:building_bler}
    \end{subfigure}
    \begin{subfigure}[b]{0.24\textwidth}
        \centering
        \includegraphics[trim=0 0 0 0, clip, width=\linewidth]{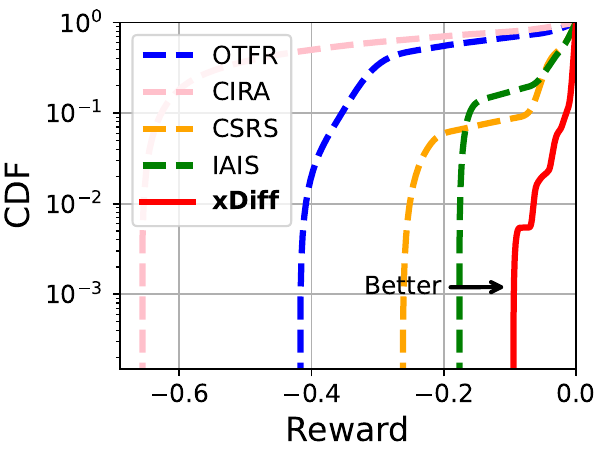}
        \caption{Reward.}
        \label{fig:building_reward}
    \end{subfigure}    
    \caption{Performance comparison of \pname and existing approaches in building scenario}
    \label{fig:compare_building}
    % \vspace{-0.15in}
\end{figure*}

\textbf{Denoising Step Number $K$.} 
The number of denoising steps is a key parameter for diffusion models, directly affecting their generation performance.  
In time-insensitive tasks such as image and video generation, a large number of denoising steps is preferable as it generally leads to better generation quality.  
However, policy-making for ICIM in O-RAN is a time-sensitive task.  
A small number of denoising steps weakens the diffusion model’s learning ability, resulting in suboptimal policies and degraded network performance.  
Conversely, a large number of denoising steps delays policy updates at DUs, leading to outdated policies that hinder resource allocation.  
To address this trade-off, we conduct experiments to empirically determine the optimal number of denoising steps that balance performance and computational delay.

Fig.~\ref{fig:ncompare} presents our experimental results.  
When increasing the number of denoising steps from 2 to 5, the network reward improves significantly due to enhanced policy quality.  
Beyond 5 steps, the reward gain diminishes; increasing the steps from 5 to 10 yields little improvement because the policy update delay at the DU offsets the benefit of improved policy quality.  
When the step number increases from 10 to 20, the reward declines.  
This occurs because the additional steps provide marginal policy enhancement while significantly prolonging computation at the Near-RT RIC, leading to outdated policies and reduced network performance.  
Based on these observations, we set the number of denoising steps to 5 in our extensive experiments.

% $K$ is a critical 
% We conduct an empirical study on the effect of the number of timesteps $K$ of our diffusion model on interference tasks. 
% Empirically we find as the N increases, the model converges faster and the performance becomes more stable. 
% Based on empirical observations in Fig. \ref{fig:ncompare}, we found that as \( K \) increases, the model converges faster and exhibits greater stability. However, this also imposes higher performance requirements on the system hardware, leading to increased training costs and a longer response time for the RIC.
% Increasing  $K$  generally results in a more stable cumulative distribution of rewards, but an excessively high $K$ may also lead to diminishing returns in performance improvement.
% In the following \pname tasks, we set a moderate value, $K = 5$, to balance the performance and computational cost. 

\textbf{Loss-Balancing Parameter \(\eta\).}  
The loss function plays a critical role in training the diffusion model, influencing both convergence speed and generation quality. 
As shown in Eq.~\eqref{eq:diff_loss_combin}, \pname{} employs a combination of two loss functions: the diffusion model’s internal loss \(\mathcal{L}_d\) and an external loss \(\mathcal{L}_q\).  
The hyperparameter \(\eta\) controls the balance between these two loss components during training.  
We conduct experiments to evaluate the impact of \(\eta\) on network performance.  

Fig.~\ref{fig:ucompare} presents our experimental results for \(\eta = 0.5\), \(1\), and \(2\).  
The results indicate that \(\eta = 1\) yields the best network performance.  
We also observed that a lower \(\eta\) (e.g., \(0.5\)) results in unstable network performance, whereas a higher \(\eta\) (e.g., \(2\)) slows down convergence.  
Based on these findings, we set \(\eta = 1\) in our extensive experiments.

% $\eta$ is a hyperparameter in the range from 0 to 2 used to balance the two loss terms during the  diffusion model training. As shown in Fig.~\ref{fig:ucompare}, the value of  $\eta$ significantly affects the performance of \pname. 
% A low value of $\eta$ (e.g.,  $\eta$ = 0.5 ) leads to an unstable performance, while a higher  $\eta$ (e.g., $\eta$ = 2) slows down convergence, as reflected in the cumulative distribution.
% Empirically, we find that setting  $\eta$ = 1  provides a favorable balance, yielding better performance in terms of reward distribution. 

\subsection{Performance Comparison}

We now evaluate the performance of \pname by comparing it against existing methods, which we describe as follows.

% In the evaluation, we use the following three ICIM methods as the comparison baselines. 

\begin{itemize}
    \item \textbf{Qualcomm's IAIS method \cite{akgun2024interference}.}  The Interference-aware Intelligent Scheduling (IAIS) method proposes a machine learning (ML)-based interference prediction technique that utilizes CSI reported by 5G UE \cite{Qualcomm_cell, Qualcomm_ue}, enabling it to schedule transmissions over dynamic air interfaces intelligently.

    \item \textbf{CSRS method \cite{wu2022demand}: }    
    This is a Cellular Spectrum Resource Sharing (CSRS) algorithm that dynamically allocates spectrum resources in cellular networks using a cluster-based approach. Center UEs share spectrum, while edge UEs use different channels. The resource allocated to a UE is proportional to its demand to minimize ICI.
    
    % By achieving higher spectral efficiency and reducing latency through fewer retransmissions.
    % \item \textbf{\pname (Our method)} 
    \item \textbf{Cell-Independent Resource Allocation (CIRA):} 
    % Each DU applies its own PF algorithm to schedule resources for UEs independently, without considering inter-cell interference. 
    This is a widely-used approach where no inter-cell coordination exists for interference management. Instead, each DU allocates the entire set of available RBs to its UEs using the PF scheduler.
    %, ensuring fair resource distribution within its coverage area, without considering ICI. 

    \item \textbf{One-Third Frequency Reuse (OTFR):} 
    In this method, the frequency spectrum is allocated equally among three DU-cells, with each cell receiving one-third of the total spectrum. This fixed allocation ensures that each cell has a dedicated portion of the spectrum, with minimal interference between them.

    % Spectrum allocation is proportional to user demand, ensuring efficient utilization and meeting QoS requirements. 
    % Integrated with the Open C-RAN architecture, the algorithm leverages centralized control for real-time adjustments.

\end{itemize}

% 由于距离太近，这个场景的全部UE都在干扰范围以内，每个gnb的最大吞吐量不到50

% used for evaluating xDiff, illustrating the deployment of multiple gNBs and the testbed arrangement. The experimental environment consists of three DU-RU: X310, N310, and B210, each positioned approximately 6 meters apart to test the realistic network conditions. The testbed (see Fig. \ref{fig:testbed}), centrally located, serves as the primary processing unit for managing data transmission and scheduling.
% %, including RIC, 5G Core, CU, GPSDO Clock and Network Switch. 
% The layout layout is intended to demonstrate indoor wireless communication scenarios. Obstacles such as office partitions, walls, etc. introduce signal attenuation, multipath and interference. This setup allows for comprehensive testing of xDiff's performance in handling dynamic resource allocation and interference mitigation in a controlled but practical environment.

\textbf{Lab-Scale Strong Interference Scenario.}
Fig.~\ref{fig:lab_senerio} shows the testbed setup in a lab scenario for this evaluation.  
This is a scenario for typical indoor 5G small cells, rich in obstacles, reflectors, blockage, human routing activities, and UE mobility. 
The experiments involve three cells and a total of ten smartphones. 
During the experiments, we use \verb|iperf| to generate time-varying data traffic for each smartphone, ranging from light traffic (35~Mbps) to heavy traffic (165~Mbps). 
We repeated the same experiments for the five ICIM methods in the same settings and collected data for 5 hours.

Fig.~\ref{fig:compare_lab} plots the throughput, queueing delay, BLER, and reward performance of the five ICIM methods.  
\textbf{Throughput:}  
It can be observed that \pname and CSRS achieve similar performance, both successfully meet the throughput demands of the UEs over time. In contrast, the throughput of the other three methods falls behind \pname and CSRS, indicating that they fail to meet the UEs' throughput demands over time.  
\textbf{Queueing Delay:}  
The results show that \pname achieves superior delay performance compared to the other four methods. This can be attributed to the fast adaptation of the diffusion model under dynamic network conditions.  
\textbf{BLER:} 
It can be seen that CIRA achieves the highest BLER, while OTFR achieves the lowest BLER. \pname yields moderate BLER compared to other methods. We note that a low BLER does not necessarily indicate better performance from a networking perspective. Instead, it may indicate that the transmission is not aggressive enough to explore the channel capacity via MCS adaptation.  
\textbf{Reward:}  
While throughput, delay, and BLER were measured during the experiments, the reward was calculated based on the measured throughput and delay values. It can be observed that \pname considerably outperforms CSRS and is significantly better than the other three methods in terms of reward. Again, this is attributed to the strong adaptability and robustness of the diffusion model in \pname in dynamic networks under time-varying interference conditions.

\textbf{Building-Scale Light Interference Scenario.}
We increase the cell radius by placing RUs at different locations in a building.
The distance between two RUs is about 18~m, with thick concrete walls in between. 
We use the existing Wi-Fi network to establish the E2 interface between Near-RT RIC and DUs, with a communication delay of about 200~ms. 
Compared to the lab scenario, this setting has a lighter interference due to the larger cell size. 
We repeated the same experiments from the previous scenario to measure the performance of the five ICIM methods.

Fig.~\ref{fig:building_reward} plots the measured per-UE throughput, queueing delay, BLER, and the calculated reward, based on which we have the following observations.  
\textbf{Throughput:}  
The experimental results show that \pname offers better throughput performance compared to the other four ICIM methods. However, unlike the previous scenario, the throughput gain of \pname is marginal. This could be attributed to the light interference in the building-scale scenario.  
\textbf{Queueing Delay:}  
\pname offers the minimum delay compared to the other four ICIM methods. This observation is consistent with that in the previous scenario.  
\textbf{BLER:}  
\pname has a moderate BLER compared to the other methods. This observation is also consistent with the previous scenario.  
\textbf{Reward:}  
\pname considerably outperforms IAIS and significantly outperforms the other three methods. This confirms that \pname is superior to the SOTA methods in both strong and light inter-cell interference networks.

% % Each gnb有两个UE在干扰范围中，一个不在干扰范围，gnb1多一个UE不在干扰范围，那么每个gnb都能达到最大100Mbps的throughput
% For building scenario, we use WI-FI to connect hosts, so RIC will use 100-200 ms to collect data and send control data via E2 Interface.

% Fig. \ref{fig:building_senerio} presents the building setup, consists of three distinct DU-RU as well, each positioned approximately 18 meters apart to test the realistic network conditions. 
%The testbed (see Fig. \ref{fig:testbed}), centrally located, serves as the primary processing unit for managing data transmission and scheduling, including RIC, 5G Core, CU, GPSDO Clock and Network Switch.

% Fig. \ref{fig:compare_building} shows a comparative evaluation of the performance of \pname with existing methods in a building scene. Subfigure (\ref{fig:building_reward}) depicts the CDF, showing that the reward performance of \pname (red dashed line) is improved and closer to zero compared to other methods. Subfigure (\ref{fig:building_thp}) depicts the CDF of throughput, showing that \pname outperforms the other methods in achieving higher throughput values. Sub-figure (\ref{fig:building_latency}) depicts the CDF of latency, in particular \pname achieves lower latency compared to Individual and Split methods. \pname's BLER (0.043) is significantly lower than that of the Individual method (0.164), indicating an improvement in reliability, and also compares favourably with the CSRS (0.053), IAIS (0.065) and Split (0.036).

\textbf{Comparison of Computational Complexity:}
We now compare the computational complexity and inference time (i.e., policy generation time) of \pname against the other four methods. Table~\ref{tab:complexity_comparison} presents our results, where the inference time was measured on a desktop with a 14th gen i9 CPU.  
Since CIRA and OTFR are rule-based scheduling methods, they offer low computational complexity and can be completed within one subframe.  
CSRS demonstrates moderate complexity with an average inference time of 5.4 ms, striking a good balance between performance and computational complexity.  
IAIS relies on machine learning for interference prediction. It has high computational complexity with an observed inference time of 38.3 ms.  
\pname has an average inference time of 21.8 ms. All these methods meet the Near-RT requirements in O-RAN systems.

% making it particularly suitable for dynamic, latency-sensitive network environments.

\begin{table}[t]
    \centering
    \caption{Comparison of computational complexity and inference time among different methods.}
    \resizebox{\columnwidth}{!}{
    \begin{tabular}{llll}    
        \toprule
        \textbf{Method} & \textbf{Computational Complexity}  & \textbf{Inference Time} \\
        \midrule
        OTFR & $\mathcal{O}(N_{\text{UEs}} \cdot N_{\text{RBs}})$ & Subframe (\(\le\)1ms)\\
        CIRA & $\mathcal{O}(1)$ & Subframe (\(\le\)1ms)\\
        CSRS \cite{wu2022demand} & $\mathcal{O}(K \cdot N_{\text{UEs}} \cdot \log N_{\text{UEs}})$ & 5.4 ms\\
        IAIS \cite{akgun2024interference} & $\mathcal{O}(N_{\text{UEs}} \cdot f_{\text{ML}})$ &   38.3 ms\\
        \textbf{\pname} & $\mathcal{O}(N_{\text{Denoising Steps}} \cdot N_{\text{UEs}} \cdot N_{\text{RBs}})$  & 21.8 ms \\
        \bottomrule
        % 0.0383 0.00537
    \end{tabular}
    }
    % \hz{add references to each method.}
    \label{tab:complexity_comparison}
\end{table}

\section{Conclusion}
In this paper, we presented \pname, an online learning-based xApp for ICIM in 5G O-RANs. \pname consists of two key components: a diffusion model and an RL framework. We formulated the ICIM problem as a reward optimization problem and employed a diffusion-based RL framework for resource allocation policy generation. To address the time-scale discrepancy between the Near-RT RIC and real-time DUs, we introduced a new concept, the \textit{preference values}, as the policy representation to bridge the operations of the RIC and DUs. We implemented \pname on a 5G testbed and evaluated its performance against state-of-the-art methods. Experimental results demonstrate that \pname outperforms existing methods, highlighting the potential of diffusion models for the online optimization of O-RAN.

% a near real-time interference map generator and a real-time MAC slider. It optimizes resource allocation using a regret-based objective function, which balances weighted throughput and latency across all sessions in an O-RAN.
%
%To address the challenges associated with network dynamics, \pname employs a multi-layer GCN to extract feature embeddings from all sessions within a resource slice. This approach effectively integrates dynamic information, enhancing decision-making for resource allocation.
% To enable online training and near real-time resource allocation control, \pname leverages a conditional diffusion policy within an online reinforcement learning framework, effectively minimizing performance regret. The diffusion policy models the stochastic nature of interference dynamics, allowing for more robust decision-making under uncertainty. By learning an adaptive policy through iterative noise sampling and refinement, \pname ensures efficient resource adaptation to varying traffic and interference conditions.
% We implemented \pname on a 5G O-RAN testbed with three RUs and evaluated its performance in two realistic scenarios. Experimental results demonstrate that \pname effectively mitigates inter-cell interference and enhances overall network efficiency.
%
%Experimental results demonstrate that \pname can reduce performance regret by 67\% compared to the state of the art.

% \clearpage

% \bibliographystyle{ACM-Reference-Format}
% \balance
% \bibliography{8_references}
\bibliographystyle{ieeetr}
% \bibliography{8_references}

\flushend
% \section{Appendix}
% \input{10_preliminary}

\end{document}